\documentclass[10pt,twocolumn,letterpaper]{article}

\usepackage{cvpr}              
\usepackage{physics}
\usepackage{multirow}
\definecolor{cvprblue}{rgb}{0.21,0.49,0.74}
\usepackage[pagebackref,breaklinks,colorlinks,allcolors=cvprblue]{hyperref}

\def\paperID{XXXX} 
\def\confName{CVPR}
\def\confYear{2026}

\title{LenghuSky-8: An 8-Year All-Sky Cloud Dataset
with Star-Aware Masks and Alt-Az Calibration
for Segmentation and Nowcasting}

\author{
Yicheng Rui$^{1,\dagger}$ \quad
Xiao-Wei Duan$^{1}$ \quad
Licai Deng$^{2}$ \quad
Fan Yang$^{2}$ \quad
Zhengming Dang$^{1}$ \quad
Zhengjun Du$^{3}$ \\
Junhao Peng$^{3}$ \quad
Wenhao Chu$^{3}$ \quad
Umut Mahmut$^{1}$ \quad
Kexin Li$^{1}$ \quad
Yiyun Wu$^{1}$ \quad
Fabo Feng$^{1}$ \\
$^{1}$State Key Laboratory of Dark Matter Physics, Tsung-Dao Lee Institute \& School of Physics and
Astronomy, \\Shanghai Jiao Tong University, Shanghai 201210, China \\
$^{2}$	National Astronomical Observatories, Chinese Academy of Sciences, Beijing 100101, China; \\
$^{3}$School of Computer Technology and Application, Qinghai University, Xining 810016, China; \\
{\tt\small ruiyicheng@sjtu.edu.cn}
}

\begin{document}
\maketitle
\begin{abstract}

Ground-based time-domain observatories require minute-by-minute, site-scale awareness of cloud cover, yet existing all-sky datasets are short, daylight-biased, or lack astrometric calibration. We present \textbf{LenghuSky-8}, an eight-year (2018–2025) all-sky imaging dataset from a premier astronomical site, comprising 429{,}620 $512{\times}512$ frames with 81.2\% night-time coverage, star-aware cloud masks, background masks, and per-pixel altitude–azimuth (alt–az) calibration. 
For robust cloud segmentation across day, night, and lunar phases, we train a linear probe on DINOv3 local features and obtain $93.3\%\pm1.1\%$ overall accuracy on a balanced, manually labeled set of 1{,}111 images. Using stellar astrometry, we map each pixel to local alt–az coordinates and measure calibration uncertainties of $\approx 0.37^\circ$ at zenith and $\approx 1.34^\circ$ at $30^\circ$ altitude, sufficient for integration with telescope schedulers.
Beyond segmentation, we introduce a short-horizon nowcasting benchmark over per-pixel three-class logits (sky/cloud/contamination) with four baselines: persistence (copying the last frame), optical flow, ConvLSTM, and VideoGPT. ConvLSTM performs best but yields only limited gains over persistence, underscoring the difficulty of near-term cloud evolution.
We release the dataset, calibrations, and an open-source toolkit for loading, evaluation, and scheduler-ready alt–az maps to boost research in segmentation, nowcasting, and autonomous observatory operations.

\end{abstract}    
\section{Introduction}
\label{sec:intro}
\begin{figure*}[!h]
    \centering
    \includegraphics[width=0.8\linewidth]{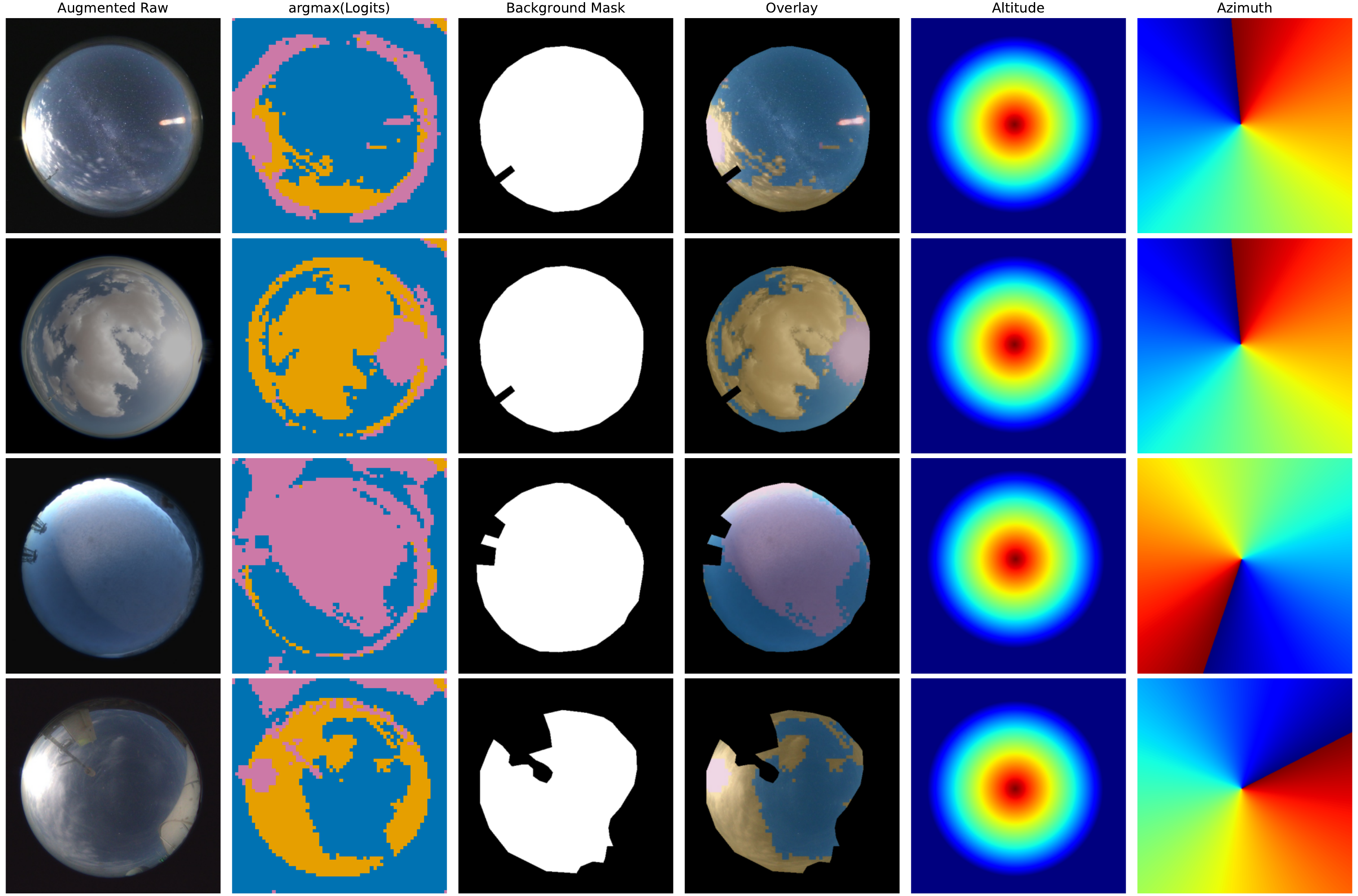}
    \caption{Samples from the dataset. First column represents the raw images that is augmented for cloud segmentation; Second column is the annotation of clear sky (blue), cloudy region (orange) and contamination region (pink); Third column is the mask for background; Forth column is the overlay of the first three columns; Last two columns are the astrometric calibration results for altitude and azimuth of the image.}
    \label{fig:overview}
\end{figure*}

Ground-based time-domain surveys such as the Zwicky Transient Facility (ZTF)~\cite{Bellm19}, the Vera C. Rubin Observatory~\cite{Ivezi19}, and the Tianyu Project~\cite{Feng24} are transforming our understanding of the dynamic Universe. To maximize scientific return, these facilities rely on cloud-aware schedulers that continuously adapt pointing and exposure plans to local, rapidly evolving weather. Achieving this requires a realistic, high-resolution cloud model that ingests recent observations and supports short-horizon predictions suitable for real-time decision making~\cite{Rui25}. Building cloud models with such granularity demands long-duration, all-sky imaging datasets collected at fixed sites, with accurate geometric calibration and night-time cloud annotations.


However, existing all-sky fisheye datasets exhibit at least one of the following limitations: (i) short temporal coverage (months rather than years), preventing seasonal modeling; (ii)  manual masking that does not scale; (iii) daytime bias, limiting astronomy use; (iv) missing astrometric calibration, making it difficult to map pixels to altitude–azimuth coordinates for effective telescope scheduling; (v) selective to easy-to-mark images that cannot depict the complicated scenario in the real world.

In this work, we introduce LenghuSky-8, an eight-year, all-sky cloud dataset with star-aware masks and alt–az calibration, collected at Lenghu, Qinghai, China—a premier astronomical site~\cite{Deng2021Natur.596..353D}—spanning 2018–2025. The dataset contains 429,620 images at resolution of 512$\times$512, covering  nights/days across 8 years, with 81.2\% night-time frames. For segmentation, we use DINOv3 local features with a linear probe, enabling robust, label-efficient separation of cloud and sky under diverse illumination, including moonlit conditions, at overall accuracy of $0.933_{-0.011}^{+0.011}$ on a small manually annotated dataset containing 1,111 images. Night-time star fields are used for astrometric calibration, yielding per-pixel altitude–azimuth coordinates with the uncertainty of $0.37^\circ$ at zenith. Samples from the dataset are shown in Fig. \ref{fig:overview}.

Our contributions are twofold:

\begin{enumerate}
    \item Dataset: an eight-year, all-sky, day–night imaging dataset with star-aware cloud masks, background annotation and alt–az calibration, suitable for both cloud nowcasting and domain-specific pretraining model; a manually labeled dataset containing 1,111 images that can be used to evaluate cloud segmentation algorithms.
    \item  Tools: a DINOv3 linear-probe segmenter for all-sky camera, a all-sky camera calibrator based on star fields, and an open evaluation toolkit with loaders, calibration maps, and scripts.
\end{enumerate}

\begin{figure*}
\centering
    \includegraphics[width = 0.8\linewidth]{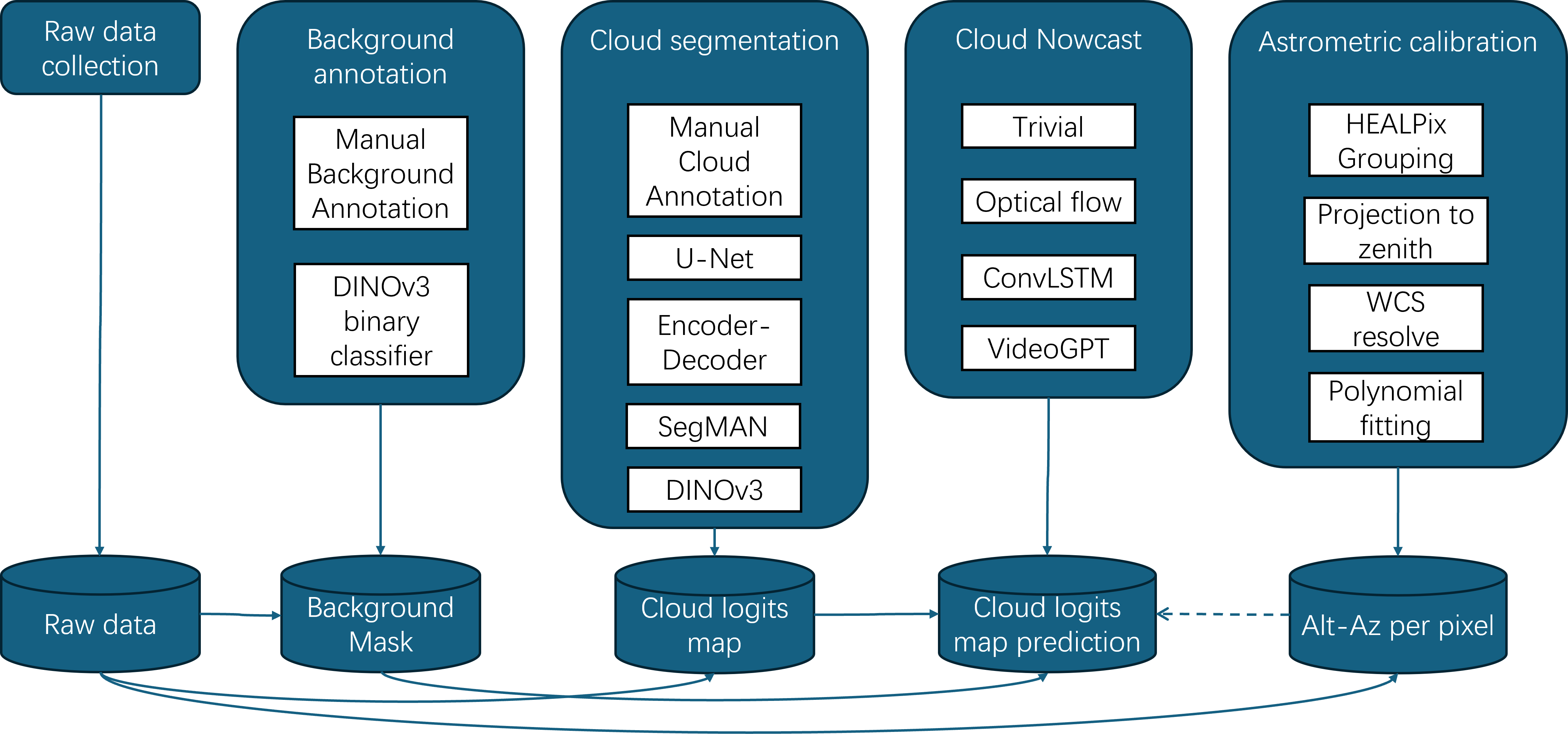}
    \caption{Workflow of this paper. Solid arrows denote dependencies among product data; dashed arrows denote potential dependencies not considered in our experiments.\label{fig:workflow}}
\end{figure*}

Workflow of this paper is shown in Fig.~\ref{fig:workflow}. Code, data, and documentation are available in \url{https://github.com/ruiyicheng/LenghuSky-8}. The remainder of the paper covers related work (Section \ref{sec:relatedwork}), dataset establishment (Section \ref{sec:dataset}), benchmark and baseline (Section \ref{sec:benchmarkandbaseline}), conclusion and discussion (Section \ref{sec:conanddis}).

\section{Related Work\label{sec:relatedwork}}
\subsection{All-sky cloud datasets}
Despite growing adoption, publicly available all-sky fisheye image datasets remain limited in both duration and scope. For example, Dev \etal  introduced the SWIMSEG daytime and nighttime cloud segmentation databases, comprising approximately 1,000 and 100 images respectively, captured in the tropical urban region of Singapore \cite{Dev17a, Dev17b}. These small-scale collections lack sufficient temporal coverage to represent seasonal or interannual cloud variability. Similarly, Li \etal  \cite{Li22} collected about 5,000 images with an all-sky camera during a site-testing campaign for the Thirty Meter Telescope (TMT) in Xinjiang. In 2019, Dev \etal  released SWINySEG, a dataset of 6,768 daytime and nighttime images annotated by human experts \cite{Dev19}. More recently, the Eye2Sky dataset \cite{schmidt2025} provided continuous all-sky imagery from 11 stations in northwestern Germany, including one site with observations spanning from April 2022 to March 2023. Nevertheless, even these larger efforts generally cover only a few months to a year and often omit detailed per-pixel cloud masks or precise geometric calibration. Consequently, the field still lacks large-scale, long-term, and well-annotated all-sky datasets necessary for robust modeling and generalizable cloud characterization.

\subsection{Cloud segmentation in all-sky camera images\label{sec:related_cloudseg}}
Ground-based all-sky cameras have been studied for cloud segmentation for over two decades. Early work relied on simple color heuristics that exploit the different scattering behavior of air molecules and cloud droplets. Fixed or adaptive thresholds on red–blue ratios, their normalized variants, and saturation/difference cues were widely used to separate cloud from clear sky in daytime scenes \cite{Long06,Ghonima12,Li11}. These methods are attractive for their robustness and real-time efficiency but can be sensitive to camera calibration, aerosol load, circumsolar saturation, thin clouds near boundaries, and star field.

Learning-based approaches reduced the need for hand-tuned thresholds by modeling sky/cloud appearance across color spaces. Dev \etal  introduced a supervised framework using partial least squares, which catalyzed reproducible evaluation across cameras and conditions~\cite{Dev17a}. Deep convolutional networks now dominate all-sky cloud segmentation. Encoder–decoder architectures (e.g., CloudSegNet) improved accuracy, especially in challenging regions near the Sun and horizon~\cite{Dev19}. U-Net variants tailored to sky imagery (CloudU-Net and SegCloud) extended segmentation across the full day and night using specialized attention modules and training on mixed day/night corpora such as SWINySEG~\cite{Shi21,Xie20}. General-purpose image segmentation archiecture like SegMAN~\cite{Fu24} are also feasible for cloud segmentation tasks.

In recent years, large-scale self-supervised pre-training has become a dominant paradigm for visual representation learning, offering powerful features that generalize across diverse visual domains. Methods such as MAE~\cite{He21}, MoCo~v3~\cite{Chen21}, iBOT~\cite{Zhou21}, and DINOv2~\cite{Oquab23} have demonstrated strong transferability to downstream tasks ranging from semantic segmentation to fine-grained recognition. These models exploit masked image modeling, contrastive learning, or teacher–student distillation to produce representations that capture both global semantics and local spatial structure. Building on these foundations, DINOv3~\cite{Oriane25} introduces improved patch-level alignment and scalable ViT backbones, enabling state-of-the-art performance on dense prediction tasks.


\subsection{Short-Term Cloud Forecasting}

Ground-based all-sky cameras are widely used to nowcast cloud fields at site scale, typically within a 5--15\,min horizon that is most relevant for robotic observatories. Early approaches advected segmented cloud masks using optical flow to extrapolate motion, demonstrating useful skill up to about 5\,min in tropical convection~\cite{Dev16}.  Hamill \etal ~\cite{Hamill93} use the cross-correlation of optical-flow based method to generate nowcast result.
Learning-based frame-to-frame warping and sky-image prediction further improved short-horizon forecasts by jointly modeling motion and deformation~\cite{Julian24}.

Beyond optical-flow extrapolation, recurrent convolutional architectures such as ConvLSTM~\cite{Shi25} formulate nowcasting as spatiotemporal sequence prediction by replacing fully connected gates with convolutions. In parallel, discrete-latent generative models such as VideoGPT—combining VQ-VAE encoders with Transformer decoders—autoregress over video tokens and offer a flexible path to learn cloud evolution priors directly from all-sky image sequences~\cite{Yan21}.

\section{Dataset\label{sec:dataset}}
\subsection{Raw data collection}
The all-sky camera is installed at the Lenghu site (longitude=$93.8961^{\circ}$, latitude=$38.6068^{\circ}$), which is located on a local summit of the Saishiteng Mountain in Qinghai, China. The altitudes of the potential observing sites range from 4,200 m to 4,500 m. In contrast, the surrounding 100,000 km$^2$ area near Lenghu Town lies at a relatively lower elevation (below 3,000 m). Both during the day and at night, the site experiences an extremely dry climate and predominantly clear skies. Such stable and arid atmospheric conditions lead to excellent seeing and low precipitable water vapor, making it an ideal site for astronomical observations.


The photographs were taken using fisheye-lens cameras. A fisheye lens is a specialized optical component designed to capture extremely wide fields of view, typically around ${180}^{\circ}$. Such lenses introduce strong visual distortion, producing wide panoramic or hemispherical images. In this work, we use a Sigma 4.5 mm f/2.8 fisheye lens, mounted on Canon 600D, 750D, and 800D all-sky camera bodies, producing raw images with resolution of $4000\times 6000$. 

The dataset can be divided into two distinct parts: data collected before 27th September 2023 18:09:48 (Part~I) and data collected thereafter (Part~II). Part~I contains images with fewer obstructions from surrounding structures or background objects, but the optical surfaces were poorly maintained, leading to a large proportion of frames affected by mud or dew on the lens. In contrast, Part~II benefits from frequent manual cleaning, which significantly improves image clarity, yet the field of view is often partially blocked by nearby objects. An example representative of Part~II is shown in the fourth row of Fig.~\ref{fig:overview}. 

\begin{figure}
    \centering
    \includegraphics[width=\linewidth]{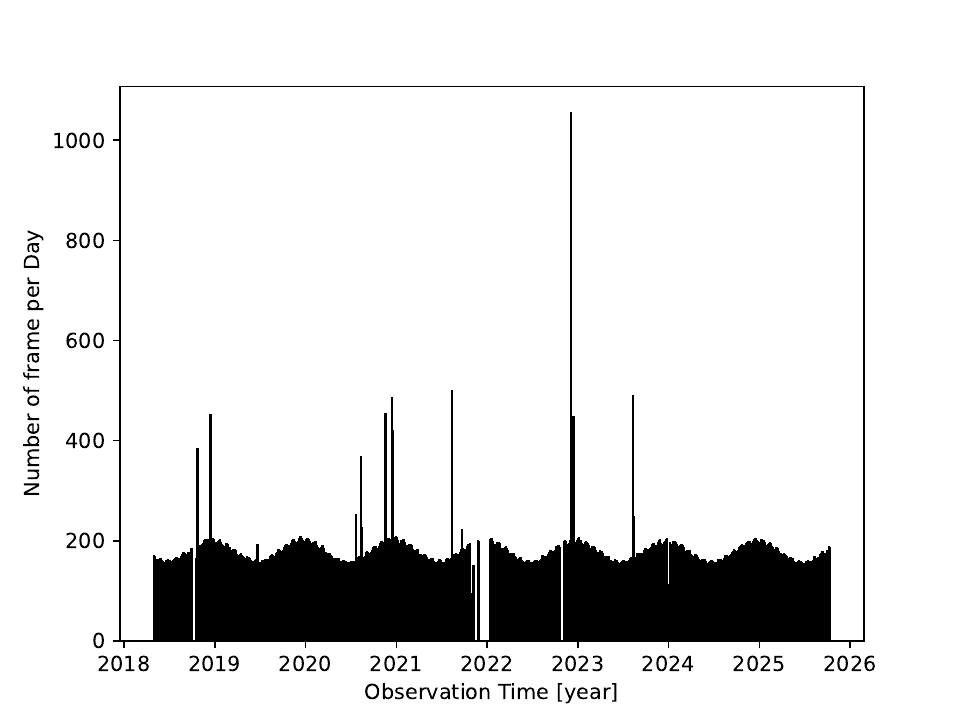}
    \caption{Daily number of captured frames in the dataset.}
    \label{fig:allstat}
\end{figure}

The distribution of the number of frames captured per day is illustrated in Fig.~\ref{fig:allstat}. The capture interval is set to 5~minutes during nighttime and 20~minutes during daytime, determined by the solar elevation angle. Consequently, the average number of frames per day is approximately 150~frames in summer and 200~frames in winter. The exposure time is dynamically adjusted according to ambient brightness, resulting in more images being captured during full-moon nights and fewer during new-moon periods. This introduces a noticeable monthly fluctuation in the total number of frames. In addition, the camera cadence is manually shortened during meteor shower events or for system testing purposes.

Continuous coverage is particularly important for tracking the evolution of cloud structures, evaluating diurnal variations, and calibrating moonlight scattering models. To ensure hour-level persistence in observations, a threshold of 60~minutes is adopted to mark a break in observation sequences. A statistical summary of the persistence characteristics is provided in Table~\ref{tab:summary_persis}. Among all observation periods, we identify 16~instances that maintain hour-level continuity for more than one month. These long-duration sequences are particularly valuable for modeling phase-dependent lunar illumination effects and evaluating temporal trends in sky conditions. The detailed list of these persistent samples is presented in Table~\ref{tab:1month}.

\begin{table*}[]
    \caption{Summary of persistence statistics for continuous observations.}
    \label{tab:summary_persis}
    \centering
    \begin{tabular}{l l l l l}
    \hline
    Duration of persistence [days] & Total time (days) & Total number of frames & Proportion of time & Proportion of frames \\
    \hline
    5 & 2176.20 & 388586 & 0.7996 & 0.9045 \\
    10 & 1540.03 & 273491 & 0.5658 & 0.6366 \\
    30 & 1000.95 & 179256 & 0.3678 & 0.4172 \\
    60 & 625.85 & 117792 & 0.2300 & 0.2742 \\
    \hline
    \end{tabular}
\end{table*}

\begin{table*}[]
    \caption{Observations with hour-level persistence for more than 1 month}
    \label{tab:1month}
    \centering
    \begin{tabular}{l l r r}
\hline
Start & End & Duration (days) & Num Frames \\
\hline
2018-12-10 13:32:44 & 2019-04-19 10:02:24 & 129.85 & 24747 \\
2020-12-06 07:22:28 & 2021-03-04 01:50:38 & 87.77 & 18034 \\
2020-01-26 19:52:42 & 2020-04-12 06:07:00 & 76.43 & 14002 \\
2021-05-31 22:22:44 & 2021-08-14 16:32:27 & 74.76 & 12206 \\
2023-01-20 06:53:42 & 2023-03-27 16:43:51 & 66.41 & 12267 \\
2019-11-20 17:12:31 & 2020-01-25 17:26:05 & 66.01 & 13083 \\
2019-08-20 21:29:09 & 2019-10-22 14:16:30 & 62.70 & 11180 \\
2020-10-03 21:41:58 & 2020-12-04 19:52:38 & 61.92 & 12273 \\
2018-05-01 00:02:44 & 2018-06-27 03:55:38 & 57.16 & 9152 \\
2021-04-04 19:40:00 & 2021-05-30 15:43:07 & 55.84 & 9221 \\
2022-07-25 22:04:45 & 2022-09-18 23:40:07 & 55.07 & 9200 \\
2020-05-10 11:27:52 & 2020-06-27 05:17:47 & 47.74 & 7514 \\
2018-08-03 05:21:52 & 2018-09-18 17:11:49 & 46.49 & 7788 \\
2023-03-28 15:04:24 & 2023-05-08 09:45:02 & 40.78 & 6930 \\
2022-04-17 16:52:51 & 2022-05-26 06:01:39 & 38.55 & 6312 \\
2018-06-30 16:15:53 & 2018-08-03 03:35:23 & 33.47 & 5347 \\
\hline
\end{tabular}

\end{table*}

\subsection{Segmentation using DINOv3\label{sec:seg}}
For cloud segmentation, we resize the center part of raw image into the resolution of $512\times 512$, normalize the image value using the range of [mean-std,mean+3$\times$std] for making the cloud more significant for annotation. Examples of the pre-processing results are shown in the first column of Fig. \ref{fig:overview}.


As clouds are amorphous objects without well-defined boundaries, constructing a reliable dataset for cloud detection poses a significant challenge. To ensure labeling accuracy, only regions with high confidence are annotated. In this work, we consider three categories: \textit{cloud}, \textit{sky}, and \textit{contamination}. Because telescopes are highly sensitive to even thin clouds, all regions unsuitable for astronomical observation are labeled as ``cloud,'' while regions that are clearly transparent are labeled as ``sky.'' Regions where classification is ambiguous—such as those covered by snow, affected by dew, saturated by sunlight, moonlight, or artificial light sources, or obscured by scattered dust—are labeled as ``contamination.'' An example of such cases is illustrated in the third row of Fig.~\ref{fig:overview}.  

We annotate a total of 1,111 images using LabelMe~\cite{Kentaro21}. The dataset is evenly distributed across moon phases (new moon, first quarter, full moon, and last quarter), times of day (02:00, 06:00, 10:00, 14:00, 18:00, and 22:00 UTC+8, which is approximately two hours ahead of local solar time), seasons (spring/autumn, summer, and winter), and cloud levels (overcast, partially cloudy, and clear). The dataset is also balanced between Part I and Part II of the observation. Examples of the annotated data are shown in Fig.~\ref{fig:annotation_segmentation} of the supplementary material. Given the inherently fuzzy and indistinct boundaries between sky, cloud, and contamination, we adopt a conservative labeling strategy, including only regions that unambiguously belong to a specific class. This approach minimizes human bias that may arise from inconsistently labeling thin cloud regions—classifying them as ``sky" in cloudy skies or as ``cloud" in clear skies.

DINOv3 (ViT-L/16) is performed for image with super-resolution to $1024\times 1024$, producing $64\times 64$ local feature vectors and $5$ global feature vectors, including 1 \textsc{CLS} and 4 register tokens. A linear probe using the local features are used for segmentation. Experiment results of using global features are  shown in Section \ref{subsec:seg-bench}. This annotator is performed on all 429,620 images.

\subsection{Background annotation}

The background remains mostly stable throughout the dataset. To improve segmentation accuracy, we manually identify frames in which the background changes. In most cases, the foreground varies while the camera remains fixed, owing to its stable mounting and the steady terrain. However, in Part~II of the dataset, a nearby building with a movable flat dome is constructed close to the all-sky camera. The frequent movement of the roof makes manual background labeling infeasible. Fortunately, the roof has two distinct and stable configurations: raised (22{:}00, full-Moon sample in Fig.~\ref{fig:annotation_segmentation} of the supplementary material) and lowered (10{:}00, first-quarter sample in Fig.~\ref{fig:annotation_segmentation} of the supplementary material). A summary of the manual annotation results is provided in Table~\ref{tab:changebkg}. The background is manually obtained for 14 distinct periods, six of which correspond to changes in the camera setup that required new astrometric solutions. The procedure for deriving these astrometric solutions is described in Section~\ref{sec:newastrometry}, and the background annotation results are given in the supplementary material.

To automatically determine the roof configuration in each image, we train a linear classifier on the DINOv3 \texttt{CLS} token using 3{,}731 images, achieving 100\% accuracy on a held-out test set of 373 images. We then combine the predictions of this classifier with the fixed manual mask to automatically annotate the background in Part~II of the dataset.

\begin{table*}[t]
  \centering
  \caption{Change of background during the observation. Background proportion is the ratio of the spherical FoV and area of the mask of background.}
  \label{tab:changebkg}
  \begin{tabular}{l l l l l l}
    \hline
    \textbf{Start time (UTC+8)} & \textbf{New astrometry} & \textbf{Roof position}& \textbf{Number of images}&\textbf{Duration [day]}&\textbf{Background proportion} \\
    \hline
    2018-05-01 00:02:44  & Yes &-- &22300&140.7&0.1917\\
    2018-09-27 19:19:49 & Yes &-- &34772&203.6&0.1357\\
    2019-04-24 15:39:36 & Yes &-- &9493&60.9&0.1702\\
    2019-06-26 18:23:18 & Yes & --&1343&8.7&0.1875\\
    2019-07-05 11:59:14  & Yes & --&70477&418.2&0.1902\\
    2020-08-26 17:24:23 & No &-- &3124&20.9&0.2093\\
    2020-09-16 15:30:23 & No & --&1178&6.9&0.2376\\
    2020-09-23 12:18:53 & No &-- &4554&25.1&0.1795\\
    2020-10-18 14:31:56 & No &-- &38258&204.0&0.2160\\
    2021-05-10 14:39:39 & No &-- &51789&380.6&0.2255\\
    2022-06-01 19:41:30  & No &-- &47589& 298.6&0.2355\\
    2023-03-27 11:15:06 & No & --&28107&184.0&0.2324\\
    2023-09-27 18:09:48 & Yes & Lowered&71976&543.9 &0.3367\\
    2023-09-27 20:47:06 & No & Raised &44660&202.0&0.3674\\
    \hline
  \end{tabular}
\end{table*}

\subsection{Astrometric calibration using stars\label{sec:newastrometry}}


\textsc{Astrometry.net} \cite{Lang10} is a widely used tool for astrometric calibration. Given star position on the image, \textsc{Astrometry.net} would return the world coordinate system (WCS), which is a reservable map from pixel space to spherical coordinate space. It utilize geometric hash matching based on the assumption of conformal transformation between detected stars and template catalogs. In general, astronomy photography have small field-of-view (FoV), which have limited geometric distortion. However, in the all-sky fisheye camera, the significant distortion would make the algorithm infeasible. A method that is similar to Jia \etal \cite{Jia25} is used mitigate this issue.  

Images are resized to resolution of $4096\times 4096$ for astrometric calibration, which corresponds to $64\times 64$ DINOv3 patches with size of $64\times 64$. Raw value of the images is preserved for star recognition. The position of stars on the image $(u,v)$ is extracted using an matched filtering algorithm, which is implemented by \textsc{sextractor}\cite{Bertin96}. Assuming the center of FoV $(u_0,v_0)$ is the origin, obtain the polar coordinate $(r,\varphi)$ in the image plane of the resolved star. Direction of the incoming light $\theta$ can be modeled by the inverse of
a type of optical projection $r = r(\theta)$ in Table \ref{tab:full_sky_proj_type}. The stars are grouped by their $(\theta,\varphi)$'s corresponding HEALPix \cite{Grosky05} patch. For each group, convert $(\theta,\varphi)$ into Cartesian coordinate $\vec{x} = (x,y,z)^T$. Rotate a group of star to $\theta=0$, which have minimal distortion, by the following matrix
\begin{equation}
\begin{aligned}
        R_z = \begin{pmatrix}\cos \varphi_c& \sin\varphi_c&0\\-\sin\varphi_c&\cos \varphi_c&0\\0&0&1\end{pmatrix};&R_y = \begin{pmatrix}\cos\theta_c&0&-\sin\theta_c\\0&1&0\\\sin\theta_c&0&\cos\theta_c\end{pmatrix};\\
         \vec{x}' &= R_yR_z\vec{x},\label{eq:rotation}
\end{aligned}
\end{equation}
where $(\theta_c,\varphi_c)$ represents the center coordinate of the HEALPix patch of this group;$\vec{x}'$ is the Cartesian coordinate system of a group of star that is on the same HEALPix patch. Projecting them back to camera plane using $r=r(\theta)$, we can obtain the undistorted star distribution. WCS of this HEALPix patch could be obtained in this way. 

When doing inference based on the results, we need first decide which HEALPix does the target land on, then do the same transformation as above. Applying the WCS on the transformed points, we could obtain the right ascension (R.A.;$\alpha$) and declination (Dec.;$\delta$) of a given pixel. Because R.A. and Dec. would change with respect time due to the Earth motion, altitude (Alt.;$h$) and azimuth (Az.;$a$) is used to present the astrometric calibration results. The transformation between $(\alpha,\delta)$ and $(h,a)$ is performed using \textsc{astropy} \cite{Astropy22}. For each time slot which requires an independent astrometric solution, as listed in Table~\ref{tab:changebkg}, we apply this procedure for multiple images. The astrometric calibration for most HEALPix patches can be obtained in this way. When obtaining the position of DINOv3 patches, we first apply the same transformation to rotate them to the center of FoV, then using the corresponding WCS to obtain the results. 

By this way, some HEALPix patches are still not resolved by \textsc{astrometry.net}, which are shown in supplementary material. We fit a radial symmetric model to interpolate the rest of the DINOv3 patches. The model is 
\begin{equation}
    \begin{aligned}
        r &= \sum_{i=0}^4 w_{2i+1} z^{2i+1}\\a &= \varphi + c,
    \end{aligned}\label{eq:fittingmodel}
\end{equation}
where $z = \frac{\pi}{2} - h$ is the zenith distance; $r$ and $\varphi$ are centered on the zenith, which is obtained by enumeration to find the highest altitude. 
 Uncertainty of the altitude $\sigma_h$ are evaluated using error propagation
\begin{equation}
    \sigma_h^2 = \sigma_z^2 = \left (\frac{\dd z}{\dd r}\right )^2\sigma_r^2. \label{eq:astrocalerrestimation}
\end{equation}

The fitting residuals are summarized in Table~\ref{tab:resastrocal}. For the camera used in this study, the assumption of an orthographic projection allows most HEALPix cells in the image to be individually resolved. The effective pixel resolution at the zenith is  $\left.\frac{\dd z}{\dd r}\right|_{z=0} \approx 3~\text{arcmin/pixel}$, which can be used to evaluate the altitude uncertainty via Eq.~\ref{eq:astrocalerrestimation}. Under the orthographic projection approximation, the average calibration uncertainties are $0.37^\circ$ in the zenith region and $1.34^\circ$ at an altitude corresponding to $z = 60^\circ$. These values are smaller than the typical field of view (FoV) of modern wide-field optical survey telescopes such as Vera C. Rubin Observatory, ZTF, and Tianyu. Detailed residual distributions from the fitting results are provided in the supplementary material.


\begin{table}[]
    \caption{Residual of astrometric calibration. $\sigma_{a}$ is the residual in the azimuth; $\sigma_{r}$ is the residual in radial, which is measured in pixel. }
    \label{tab:resastrocal}
    \centering
    \begin{tabular}{lll}\hline
       Start time (UTC+8) & $\sigma_{\text{a}}$ [deg]& $\sigma_{\text{r}}$ [pixel]\\\hline
       2018-05-01 00:02:44&0.569&4.494\\
       2018-09-27 19:19:49&1.382&9.016\\
       2019-04-24 15:39:36&0.774&4.298\\
       2019-06-26 18:23:18&0.814&5.239\\
       2019-07-05 11:59:14&1.978&5.958\\
       2023-09-27 18:09:48&2.256&15.086\\\hline
       Average&1.295&7.349\\\hline
    \end{tabular}

\end{table}

\section{Benchmark and baseline\label{sec:benchmarkandbaseline}}
\subsection{Ambiguity-Aware Sky/Cloud Segmentation Benchmark}
\label{subsec:seg-bench}
Because only confident regions are annotated manually, we only consider the metric that is on the the annotated region. 
We calculate the Accuracy, Recall, F1 score for the test set for the regions that we have annotated. Because  a conservative strategy is applied for annotation, as described in Section~\ref{sec:seg}, the classical metric for object segmentation like mIoU is not feasible for this task. Train-test-valid set are taken at the ratio of 8:1:1, i.e. 889 samples for training set, 111 samples for validation set and 111 samples for test set. For each baseline, we bootstrap the train-valid-test set separation for 20 times to determine the uncertainty of the models' performance. The following models are used as the baseline in this work.

\textbf{Linear probe of DINOv3}: As described in section \ref{sec:seg}, a linear probe of DINOv3 (ViT-L/16) model is applied on patched of the output embeddings. Besides the local embedding, DINOv3 also have 1 \texttt{CLS} token and 4 register tokens to present the global features. We considered 4 setup for the utilization of DINOv3 model: (1) local feature tokens only; (2) local feature concatenate with \texttt{CLS} tokens; (3) local feature concatenate with the mean of \texttt{CLS} + register tokens; (4) local features concatenate with \texttt{CLS} and all register tokens.

\textbf{Encoder-decoder (CloudSegNet-like)}: As mentioned in \ref{sec:related_cloudseg}, CloudSegNet is a network that is dedicated designed for cloud segmentation using encoder-decoder. In this work, an encoder-decoder is implemented as a baseline. We adapt the network for the $512\times 512$ image size. Meanwhile, the final output layer is used to output $512\times 512 \times 3$ logits to enable three-class segmentation. Details of the implementation is shown in supplementary material.

\textbf{U-Net (CloudU-Net-like)}: As a widely used image segmentation algorithm, U-Net is also implemented in this work as a baseline for cloud segmentation. We also adapted the shape of output for this network. Details of the implementation is shown in supplementary material.

\textbf{SegMAN}:  As mentioned in Section \ref{sec:related_cloudseg}, SegMAN is a state-of-the-art image segmentation architecture. We adapt the input shape for the experiment. Tiny, small, base, and large models are implemented for comparison.


A comparison of the results is shown in Table~\ref{tab:overallmetricseg}. As shown in the table, a DINOv3 linear probe using only local features achieves an overall test accuracy of $0.933_{-0.011}^{+0.011}\%$. In contrast, concatenating local features with the mean of the \texttt{[CLS]} and register tokens yields $0.934_{-0.012}^{+0.010}\%$ overall test accuracy. The two methods show no statistically significant difference on the test set. Therefore, we adopt the simpler approach—linear probing of DINOv3 with local-feature embeddings—as our cloud segmentation method. Per-class segmentation metrics for these baselines are provided in the supplementary material.

\begin{table*}[ht]
\centering
\caption{Overall metrics for cloud segmentation(median with 16th-83rd percentiles)\label{tab:overallmetricseg}}
\begin{tabular}{lcccc}
\hline
Baseline & Accuracy & Macro Precision & Macro Recall & Macro F1-score \\
\hline
DINOv3 local &  $0.933_{-0.011}^{+0.011}$ & $0.894_{-0.028}^{+0.020}$ & $0.867_{-0.037}^{+0.026}$ & $0.873_{-0.019}^{+0.028}$ \\
DINOv3 local + \texttt{CLS}&$0.930_{-0.011}^{+0.013}$ & $0.890_{-0.021}^{+0.013}$ & $0.853_{-0.029}^{+0.045}$ & $0.868_{-0.021}^{+0.022}$ \\
DINOv3 local + \texttt{CLS} + register & $0.924_{-0.010}^{+0.020}$ & $0.897_{-0.030}^{+0.008}$ & $0.841_{-0.031}^{+0.046}$ & $0.860_{-0.023}^{+0.033}$ \\
DINOv3 local + mean(\texttt{CLS} + register) & $0.934_{-0.012}^{+0.010}$ & $0.892_{-0.020}^{+0.017}$ & $0.866_{-0.036}^{+0.036}$ & $0.872_{-0.014}^{+0.020}$ \\
Encoder-decoder & $0.804_{-0.027}^{+0.014}$ & $0.771_{-0.025}^{+0.037}$ & $0.655_{-0.016}^{+0.033}$ & $0.686_{-0.033}^{+0.022}$ \\
U-Net & $0.851_{-0.020}^{+0.031}$ & $0.839_{-0.080}^{+0.030}$ & $0.699_{-0.034}^{+0.022}$ & $0.728_{-0.048}^{+0.015}$ \\
SegMAN Tiny & \textbf{$0.885_{-0.023}^{+0.015}$} & $0.801_{-0.031}^{+0.049}$ & \textbf{$0.760_{-0.026}^{+0.038}$} & \textbf{$0.772_{-0.018}^{+0.038}$} \\
SegMAN Small & $0.881_{-0.027}^{+0.013}$ & $0.812_{-0.036}^{+0.037}$ & $0.748_{-0.023}^{+0.031}$ & $0.767_{-0.020}^{+0.015}$ \\
SegMAN Base & $0.882_{-0.019}^{+0.008}$ & \textbf{$0.817_{-0.017}^{+0.025}$} & $0.735_{-0.028}^{+0.033}$ & $0.761_{-0.032}^{+0.024}$ \\
SegMAN Large & $0.872_{-0.021}^{+0.014}$ & $0.804_{-0.033}^{+0.038}$ & $0.725_{-0.017}^{+0.023}$ & $0.748_{-0.022}^{+0.016}$ \\
\hline
\end{tabular}
\end{table*}

\subsection{Weather Nowcast Benchmark}
\label{subsec:nowcast-bench}

The input to this task consists of two consecutive frames of logits generated by a DINOv3 linear probe applied to local patches. The objective is to predict the segmentation of the subsequent frame based on the preceding $n$ frames. The model provides a three-class prediction: clear, cloud, and contamination. The ground truth is derived from an inference-based tri-label map. As outlined in Section \ref{subsec:seg-bench}, only non-background pixels are considered in the scoring. Unlike direct image prediction, forecasting logits is a more meaningful approach for cloud modeling. The training dataset spans from May 1st, 2018 to January 1st, 2023, intersected with data from September 28th, 2023 to January 1st, 2025. The remaining data are used for testing. The following models are considered as baselines in this work.

\textbf{Trivial baseline}: For comparison, we setup trivial baseline: An identical map from previous frame to next frame. This baseline is expected to have moderate performance because the prediction would be correct if the whole sky is clear (sky), cloudy (cloud) or covered by snow/ice (contamination) for a long time.

\textbf{Optical Flow extrapolation}: We implement the optical flow based algorithm as introduced in Hamill et. al \cite{Hamill93} as an baseline. This algorithm predicts future frames by extrapolating motion patterns from historical data. The core algorithm uses Farneback's dense optical flow method \cite{Farneb03} to compute motion vectors between the last two input frames.  For prediction, the it assumes motion continuity by applying this computed flow field to warp the most recent frame forward, effectively propagating each pixel along its estimated trajectory. 

\textbf{ConvLSTM}: ConvLSTM treats the past 
 logit maps as a spatiotemporal tensor and uses convolutional gating to retain localized motion/morphology cues, predicting the next tri-label (sky/cloud/contamination) logits end-to-end; training and scoring follow the background masking used throughout the benchmark (only judgeable, non-background pixels contribute to the loss/metrics).

\textbf{VideoGPT}: VideoGPT-style generative baseline tokenizes each 3-channel logit map with a VQ-VAE, then applies a causal transformer to autoregress over space–time token sequences to synthesize the next frame’s tokens, which are decoded back to logits. 

Results of these baselines are shown in Table~\ref{tab:overallmetricfore}. The per-class segmentation metric for these baselines are shown in  supplementary material. It is interesting to see that ConvLSTM have highest prediction accuracy and VideoGPT model have worst performance. Meanwhile number of used past frame would not significantly affect the prediction accuracy. Details of the model implementation are shown in supplementary material.

\begin{table}[ht]
\centering
\caption{Overall metrics for weather nowcast\label{tab:overallmetricfore}. VideoGPT-$n$ refers to using VideoGPT to predict the next frame according to previous $n$ frames; Accuracy, precision, recall, and F1-score are measure by the macro average among sky, cloud and contamination class. }
\begin{tabular}{lcccc}
\hline
Baseline & Accuracy & Precision & Recall & F1-score \\
\hline
Trivial & 0.888 & 0.859 & 0.858 & 0.858 \\
Optical Flow & 0.888 & 0.858 & 0.860 & 0.859 \\
ConvLSTM & 0.890 & 0.871 & 0.849 & 0.859 \\
VideoGPT-1 & 0.871 & 0.841 & 0.823 & 0.831 \\
VideoGPT-2 & 0.872 & 0.840 & 0.825 & 0.831 \\
VideoGPT-7 & 0.870 & 0.841 & 0.818 & 0.827 \\
\hline
\end{tabular}
\end{table}

\section{Conclusion and Discussion\label{sec:conanddis}}

This work presents LenghuSky-8, an 8-year all-sky cloud dataset with star-aware masks and Alt-Az calibration for segmentation and nowcasting. This dataset have long temporal coverage, full automatic cloud annotation at $93.3\%$ median accuracy, coverage of both day time and night time, astrometric calibration to map pixels to local altitude and azimuth, and complete sample that include various moon phase and conditions.  This dataset is important for modeling local cloud environment, which is useful for developing the scheduler of automatic ground-based survey telescopes. Meanwhile, it is a valuable dataset for environmental study.

Besides, different configurations of DINOv3, encoder-decoder, U-Net, and various size of SegMAN are tested for cloud segmentation. Linear probe on local features of DINOv3 have the highest accuracy, which show the potential to adopt pre-training model to various fields. We need to mention that metrics exclude ambiguous areas are used for this task, which may inflate performance and obscure boundary mistakes. Furthermore, we test trivial baseline, optical flow extrapolation, ConvLSTM, and VideoGPT on the weather nowcast task. Interestingly, VideoGPT achieve the worst performance in these model. In future work, we would collect more annotated images at diverse observation sites and instruments to enhance the generalization of the trained model. 

In the cloud nowcasting task, optical flow extrapolation and ConvLSTM algorithm do not have significant advantage over trivial identical map from the previous frame to the next one, which is a common phenomenon for time series prediction tasks~\cite{Wang24,Zhang25}. Future research are required to propose a more accurate way for cloud nowcasting. As shown in Fig.~\ref{fig:workflow}, incorporating astrometric calibration to register frames into a common sky coordinate system may reduce spurious motion and stabilize training. Likewise, injecting physically motivated structure—e.g., advection consistency, non-negativity and boundedness of radiometric quantities, or weak mass-continuity priors—can regularize solutions and encourage physically plausible evolution.

\section*{Acknowledgement}

This work is supported by the National Key R\&D Program of China, Nos. 2024YFA1611801 and 2024YFC2207700, the National Natural Science Foundation of China (NSFC) under grant No.12473066, No.12233009 and No. 62562052,
the Shanghai Jiao Tong University 2030 Initiative, the Basic Resources Investigation Program of the Ministry of Science and Technology of China (Grant No. 2023FY101100) and Yuanqi Observatory. We also sincerely thank Yansong Wang, Jinfang Zhang, Zixuan Yang, Baolong Ma, Qin Ma, and Xiangyu Ma for their valuable work in data labeling.

\newpage
{
    \small
    \bibliographystyle{ieeenat_fullname}
    \bibliography{main}

@ARTICLE{Bellm19,
       author = {{Bellm}, Eric C. and {Kulkarni}, Shrinivas R. and {Graham}, Matthew J. and {Dekany}, Richard and {Smith}, Roger M. and {Riddle}, Reed and {Masci}, Frank J. and {Helou}, George and {Prince}, Thomas A. and {Adams}, Scott M. and {Barbarino}, C. and {Barlow}, Tom and {Bauer}, James and {Beck}, Ron and {Belicki}, Justin and {Biswas}, Rahul and {Blagorodnova}, Nadejda and {Bodewits}, Dennis and {Bolin}, Bryce and {Brinnel}, Valery and {Brooke}, Tim and {Bue}, Brian and {Bulla}, Mattia and {Burruss}, Rick and {Cenko}, S. Bradley and {Chang}, Chan-Kao and {Connolly}, Andrew and {Coughlin}, Michael and {Cromer}, John and {Cunningham}, Virginia and {De}, Kishalay and {Delacroix}, Alex and {Desai}, Vandana and {Duev}, Dmitry A. and {Eadie}, Gwendolyn and {Farnham}, Tony L. and {Feeney}, Michael and {Feindt}, Ulrich and {Flynn}, David and {Franckowiak}, Anna and {Frederick}, S. and {Fremling}, C. and {Gal-Yam}, Avishay and {Gezari}, Suvi and {Giomi}, Matteo and {Goldstein}, Daniel A. and {Golkhou}, V. Zach and {Goobar}, Ariel and {Groom}, Steven and {Hacopians}, Eugean and {Hale}, David and {Henning}, John and {Ho}, Anna Y.~Q. and {Hover}, David and {Howell}, Justin and {Hung}, Tiara and {Huppenkothen}, Daniela and {Imel}, David and {Ip}, Wing-Huen and {Ivezi{\'c}}, {\v{Z}}eljko and {Jackson}, Edward and {Jones}, Lynne and {Juric}, Mario and {Kasliwal}, Mansi M. and {Kaspi}, S. and {Kaye}, Stephen and {Kelley}, Michael S.~P. and {Kowalski}, Marek and {Kramer}, Emily and {Kupfer}, Thomas and {Landry}, Walter and {Laher}, Russ R. and {Lee}, Chien-De and {Lin}, Hsing Wen and {Lin}, Zhong-Yi and {Lunnan}, Ragnhild and {Giomi}, Matteo and {Mahabal}, Ashish and {Mao}, Peter and {Miller}, Adam A. and {Monkewitz}, Serge and {Murphy}, Patrick and {Ngeow}, Chow-Choong and {Nordin}, Jakob and {Nugent}, Peter and {Ofek}, Eran and {Patterson}, Maria T. and {Penprase}, Bryan and {Porter}, Michael and {Rauch}, Ludwig and {Rebbapragada}, Umaa and {Reiley}, Dan and {Rigault}, Mickael and {Rodriguez}, Hector and {van Roestel}, Jan and {Rusholme}, Ben and {van Santen}, Jakob and {Schulze}, S. and {Shupe}, David L. and {Singer}, Leo P. and {Soumagnac}, Maayane T. and {Stein}, Robert and {Surace}, Jason and {Sollerman}, Jesper and {Szkody}, Paula and {Taddia}, F. and {Terek}, Scott and {Van Sistine}, Angela and {van Velzen}, Sjoert and {Vestrand}, W. Thomas and {Walters}, Richard and {Ward}, Charlotte and {Ye}, Quan-Zhi and {Yu}, Po-Chieh and {Yan}, Lin and {Zolkower}, Jeffry},
        title = "{The Zwicky Transient Facility: System Overview, Performance, and First Results}",
      journal = {Publications of the Astronomical Society of the Pacific},
         year = 2019,
        month = jan,
       volume = {131},
       number = {995},
        pages = {018002},
          doi = {10.1088/1538-3873/aaecbe},
archivePrefix = {arXiv},
       eprint = {1902.01932},
 primaryClass = {astro-ph.IM},
       adsurl = {https://ui.adsabs.harvard.edu/abs/2019PASP..131a8002B}
}

@ARTICLE{Ivezi19,
       author = {{Ivezi{\'c}}, {\v{Z}}eljko and {Kahn}, Steven M. and {Tyson}, J. Anthony and {Abel}, Bob and {Acosta}, Emily and {Allsman}, Robyn and {Alonso}, David and {AlSayyad}, Yusra and {Anderson}, Scott F. and {Andrew}, John and {Angel}, James Roger P. and {Angeli}, George Z. and {Ansari}, Reza and {Antilogus}, Pierre and {Araujo}, Constanza and {Armstrong}, Robert and {Arndt}, Kirk T. and {Astier}, Pierre and {Aubourg}, {\'E}ric and {Auza}, Nicole and {Axelrod}, Tim S. and {Bard}, Deborah J. and {Barr}, Jeff D. and {Barrau}, Aurelian and {Bartlett}, James G. and {Bauer}, Amanda E. and {Bauman}, Brian J. and {Baumont}, Sylvain and {Bechtol}, Ellen and {Bechtol}, Keith and {Becker}, Andrew C. and {Becla}, Jacek and {Beldica}, Cristina and {Bellavia}, Steve and {Bianco}, Federica B. and {Biswas}, Rahul and {Blanc}, Guillaume and {Blazek}, Jonathan and {Blandford}, Roger D. and {Bloom}, Josh S. and {Bogart}, Joanne and {Bond}, Tim W. and {Booth}, Michael T. and {Borgland}, Anders W. and {Borne}, Kirk and {Bosch}, James F. and {Boutigny}, Dominique and {Brackett}, Craig A. and {Bradshaw}, Andrew and {Brandt}, William Nielsen and {Brown}, Michael E. and {Bullock}, James S. and {Burchat}, Patricia and {Burke}, David L. and {Cagnoli}, Gianpietro and {Calabrese}, Daniel and {Callahan}, Shawn and {Callen}, Alice L. and {Carlin}, Jeffrey L. and {Carlson}, Erin L. and {Chandrasekharan}, Srinivasan and {Charles-Emerson}, Glenaver and {Chesley}, Steve and {Cheu}, Elliott C. and {Chiang}, Hsin-Fang and {Chiang}, James and {Chirino}, Carol and {Chow}, Derek and {Ciardi}, David R. and {Claver}, Charles F. and {Cohen-Tanugi}, Johann and {Cockrum}, Joseph J. and {Coles}, Rebecca and {Connolly}, Andrew J. and {Cook}, Kem H. and {Cooray}, Asantha and {Covey}, Kevin R. and {Cribbs}, Chris and {Cui}, Wei and {Cutri}, Roc and {Daly}, Philip N. and {Daniel}, Scott F. and {Daruich}, Felipe and {Daubard}, Guillaume and {Daues}, Greg and {Dawson}, William and {Delgado}, Francisco and {Dellapenna}, Alfred and {de Peyster}, Robert and {de Val-Borro}, Miguel and {Digel}, Seth W. and {Doherty}, Peter and {Dubois}, Richard and {Dubois-Felsmann}, Gregory P. and {Durech}, Josef and {Economou}, Frossie and {Eifler}, Tim and {Eracleous}, Michael and {Emmons}, Benjamin L. and {Fausti Neto}, Angelo and {Ferguson}, Henry and {Figueroa}, Enrique and {Fisher-Levine}, Merlin and {Focke}, Warren and {Foss}, Michael D. and {Frank}, James and {Freemon}, Michael D. and {Gangler}, Emmanuel and {Gawiser}, Eric and {Geary}, John C. and {Gee}, Perry and {Geha}, Marla and {Gessner}, Charles J.~B. and {Gibson}, Robert R. and {Gilmore}, D. Kirk and {Glanzman}, Thomas and {Glick}, William and {Goldina}, Tatiana and {Goldstein}, Daniel A. and {Goodenow}, Iain and {Graham}, Melissa L. and {Gressler}, William J. and {Gris}, Philippe and {Guy}, Leanne P. and {Guyonnet}, Augustin and {Haller}, Gunther and {Harris}, Ron and {Hascall}, Patrick A. and {Haupt}, Justine and {Hernandez}, Fabio and {Herrmann}, Sven and {Hileman}, Edward and {Hoblitt}, Joshua and {Hodgson}, John A. and {Hogan}, Craig and {Howard}, James D. and {Huang}, Dajun and {Huffer}, Michael E. and {Ingraham}, Patrick and {Innes}, Walter R. and {Jacoby}, Suzanne H. and {Jain}, Bhuvnesh and {Jammes}, Fabrice and {Jee}, M. James and {Jenness}, Tim and {Jernigan}, Garrett and {Jevremovi{\'c}}, Darko and {Johns}, Kenneth and {Johnson}, Anthony S. and {Johnson}, Margaret W.~G. and {Jones}, R. Lynne and {Juramy-Gilles}, Claire and {Juri{\'c}}, Mario and {Kalirai}, Jason S. and {Kallivayalil}, Nitya J. and {Kalmbach}, Bryce and {Kantor}, Jeffrey P. and {Karst}, Pierre and {Kasliwal}, Mansi M. and {Kelly}, Heather and {Kessler}, Richard and {Kinnison}, Veronica and {Kirkby}, David and {Knox}, Lloyd and {Kotov}, Ivan V. and {Krabbendam}, Victor L. and {Krughoff}, K. Simon and {Kub{\'a}nek}, Petr and {Kuczewski}, John and {Kulkarni}, Shri and {Ku}, John and {Kurita}, Nadine R. and {Lage}, Craig S. and {Lambert}, Ron and {Lange}, Travis and {Langton}, J. Brian and {Le Guillou}, Laurent and {Levine}, Deborah and {Liang}, Ming and {Lim}, Kian-Tat and {Lintott}, Chris J. and {Long}, Kevin E. and {Lopez}, Margaux and {Lotz}, Paul J. and {Lupton}, Robert H. and {Lust}, Nate B. and {MacArthur}, Lauren A. and {Mahabal}, Ashish and {Mandelbaum}, Rachel and {Markiewicz}, Thomas W. and {Marsh}, Darren S. and {Marshall}, Philip J. and {Marshall}, Stuart and {May}, Morgan and {McKercher}, Robert and {McQueen}, Michelle and {Meyers}, Joshua and {Migliore}, Myriam and {Miller}, Michelle and {Mills}, David J.},
        title = "{LSST: From Science Drivers to Reference Design and Anticipated Data Products}",
      journal = {The Astrophysical Journal},
         year = 2019,
        month = mar,
       volume = {873},
       number = {2},
          eid = {111},
        pages = {111},
          doi = {10.3847/1538-4357/ab042c},
archivePrefix = {arXiv},
       eprint = {0805.2366},
 primaryClass = {astro-ph},
       adsurl = {https://ui.adsabs.harvard.edu/abs/2019ApJ...873..111I}
}

@ARTICLE{Feng24,
       author = {{Feng}, F.~B. and {Rui}, Y.~C. and {Du}, Z.~M. and {Lin}, Q. and {Zhang}, C.~C. and {Zhou}, D. and {Cui}, K.~M. and {Ogihara}, M. and {Yang}, M. and {Lin}, J. and {Cai}, Y.~Z. and {Yang}, T.~Z. and {Pang}, X.~Y. and {Jian}, M.~J. and {Li}, W.~X. and {Guo}, H.~X. and {Shi}, X. and {Shi}, J.~C. and {Li}, J.~Y. and {Guo}, K.~R. and {Yao}, S. and {Chen}, A.~M. and {Jia}, P. and {Tan}, X.~Y. and {Jenkins}, S.~J. and {Jiang}, H.~X. and {Zhang}, M.~Y. and {Li}, K.~X. and {Xiao}, G.~Y. and {Zheng}, S.~Y. and {Xuan}, Y.~F. and {Zheng}, J. and {He}, M. and {Jones}, R.~A.~H. and {Song}, C.~Y.},
        title = "{Tianyu-Search for the Second Solar System and Explore the Dynamic Univers}",
      journal = {Acta Astronomica Sinica},
         year = 2024,
        month = jul,
       volume = {65},
       number = {4},
          eid = {34},
        pages = {34},
          doi = {10.15940/j.cnki.0001-5245.2024.04.001},
archivePrefix = {arXiv},
       eprint = {2404.07149},
 primaryClass = {astro-ph.IM},
       adsurl = {https://ui.adsabs.harvard.edu/abs/2024AcASn..65...34F}
}

@ARTICLE{Long06,
       author = {{Long}, C.~N. and {Sabburg}, J.~M. and {Calb{\'o}}, J. and {Pag{\`e}s}, D.},
        title = "{Retrieving Cloud Characteristics from Ground-Based Daytime Color All-Sky Images}",
      journal = {Journal of Atmospheric and Oceanic Technology},
         year = 2006,
        month = jan,
       volume = {23},
       number = {5},
        pages = {633},
          doi = {10.1175/JTECH1875.1},
       adsurl = {https://ui.adsabs.harvard.edu/abs/2006JAtOT..23..633L}
}

@Article{Ghonima12,
AUTHOR = {Ghonima, M. S. and Urquhart, B. and Chow, C. W. and Shields, J. E. and Cazorla, A. and Kleissl, J.},
TITLE = {A method for cloud detection and opacity classification based on ground based sky imagery},
JOURNAL = {Atmospheric Measurement Techniques},
VOLUME = {5},
YEAR = {2012},
NUMBER = {11},
PAGES = {2881--2892},
URL = {https://amt.copernicus.org/articles/5/2881/2012/},
DOI = {10.5194/amt-5-2881-2012}
}

@ARTICLE{Li11,
       author = {{Li}, Qingyong and {Lu}, Weitao and {Yang}, Jun},
        title = "{A Hybrid Thresholding Algorithm for Cloud Detection on Ground-Based Color Images}",
      journal = {Journal of Atmospheric and Oceanic Technology},
         year = 2011,
        month = oct,
       volume = {28},
       number = {10},
        pages = {1286-1296},
          doi = {10.1175/JTECH-D-11-00009.1},
       adsurl = {https://ui.adsabs.harvard.edu/abs/2011JAtOT..28.1286L}
}

@ARTICLE{Deng2021Natur.596..353D,
       author = {{Deng}, Licai and {Yang}, Fan and {Chen}, Xiaodian and {He}, Fei and {Liu}, Qili and {Zhang}, Bo and {Zhang}, Chunguang and {Wang}, Kun and {Liu}, Nian and {Ren}, Anbing and {Luo}, Zhiquan and {Yan}, Zhengzhou and {Tian}, Jianfeng and {Pan}, Jun},
        title = "{Lenghu on the Tibetan Plateau as an astronomical observing site}",
      journal = {Nature},
         year = 2021,
        month = aug,
       volume = {596},
       number = {7872},
        pages = {353-356},
          doi = {10.1038/s41586-021-03711-z},
       adsurl = {https://ui.adsabs.harvard.edu/abs/2021Natur.596..353D}
}

@ARTICLE{Dev19,
       author = {{Dev}, Soumyabrata and {Nautiyal}, Atul and {Lee}, Yee Hui and {Winkler}, Stefan},
        title = "{CloudSegNet: A Deep Network for Nychthemeron Cloud Image Segmentation}",
      journal = {IEEE Geoscience and Remote Sensing Letters},
         year = 2019,
        month = dec,
       volume = {16},
       number = {12},
        pages = {1814-1818},
          doi = {10.1109/LGRS.2019.2912140},
archivePrefix = {arXiv},
       eprint = {1904.07979},
 primaryClass = {physics.ao-ph},
       adsurl = {https://ui.adsabs.harvard.edu/abs/2019IGRSL..16.1814D}
}

@ARTICLE{Julian24,
       author = {{Julian}, Leron and {Sankaranarayanan}, Aswin C.},
        title = "{Precise Forecasting of Sky Images Using Spatial Warping}",
      journal = {arXiv e-prints},
         year = 2024,
        month = sep,
          eid = {arXiv:2409.12162},
        pages = {arXiv:2409.12162},
          doi = {10.48550/arXiv.2409.12162},
archivePrefix = {arXiv},
       eprint = {2409.12162},
 primaryClass = {cs.CV},
       adsurl = {https://ui.adsabs.harvard.edu/abs/2024arXiv240912162J}
}

@inproceedings{Dev16,
author = {Dev, Soumyabrata and Savoy, Florian and Lee, Yee Hui},
year = {2016},
month = {11},
pages = {2563-2566},
title = {Short-term prediction of localized cloud motion using ground-based sky imagers},
doi = {10.1109/TENCON.2016.7848499}
}

@ARTICLE{Bertin96,
       author = {{Bertin}, E. and {Arnouts}, S.},
        title = "{SExtractor: Software for source extraction.}",
      journal = {Astronomy and Astrophysics Supplement Series},
         year = 1996,
        month = jun,
       volume = {117},
        pages = {393-404},
          doi = {10.1051/aas:1996164},
       adsurl = {https://ui.adsabs.harvard.edu/abs/1996A&AS..117..393B}
}

@ARTICLE{Astropy22,
       author = {{Astropy Collaboration} and {Price-Whelan}, Adrian M. and {Lim}, Pey Lian and {Earl}, Nicholas and {Starkman}, Nathaniel and {Bradley}, Larry and {Shupe}, David L. and {Patil}, Aarya A. and {Corrales}, Lia and {Brasseur}, C.~E. and {N{\"o}the}, Maximilian and {Donath}, Axel and {Tollerud}, Erik and {Morris}, Brett M. and {Ginsburg}, Adam and {Vaher}, Eero and {Weaver}, Benjamin A. and {Tocknell}, James and {Jamieson}, William and {van Kerkwijk}, Marten H. and {Robitaille}, Thomas P. and {Merry}, Bruce and {Bachetti}, Matteo and {G{\"u}nther}, H. Moritz and {Aldcroft}, Thomas L. and {Alvarado-Montes}, Jaime A. and {Archibald}, Anne M. and {B{\'o}di}, Attila and {Bapat}, Shreyas and {Barentsen}, Geert and {Baz{\'a}n}, Juanjo and {Biswas}, Manish and {Boquien}, M{\'e}d{\'e}ric and {Burke}, D.~J. and {Cara}, Daria and {Cara}, Mihai and {Conroy}, Kyle E. and {Conseil}, Simon and {Craig}, Matthew W. and {Cross}, Robert M. and {Cruz}, Kelle L. and {D'Eugenio}, Francesco and {Dencheva}, Nadia and {Devillepoix}, Hadrien A.~R. and {Dietrich}, J{\"o}rg P. and {Eigenbrot}, Arthur Davis and {Erben}, Thomas and {Ferreira}, Leonardo and {Foreman-Mackey}, Daniel and {Fox}, Ryan and {Freij}, Nabil and {Garg}, Suyog and {Geda}, Robel and {Glattly}, Lauren and {Gondhalekar}, Yash and {Gordon}, Karl D. and {Grant}, David and {Greenfield}, Perry and {Groener}, Austen M. and {Guest}, Steve and {Gurovich}, Sebastian and {Handberg}, Rasmus and {Hart}, Akeem and {Hatfield-Dodds}, Zac and {Homeier}, Derek and {Hosseinzadeh}, Griffin and {Jenness}, Tim and {Jones}, Craig K. and {Joseph}, Prajwel and {Kalmbach}, J. Bryce and {Karamehmetoglu}, Emir and {Ka{\l}uszy{\'n}ski}, Miko{\l}aj and {Kelley}, Michael S.~P. and {Kern}, Nicholas and {Kerzendorf}, Wolfgang E. and {Koch}, Eric W. and {Kulumani}, Shankar and {Lee}, Antony and {Ly}, Chun and {Ma}, Zhiyuan and {MacBride}, Conor and {Maljaars}, Jakob M. and {Muna}, Demitri and {Murphy}, N.~A. and {Norman}, Henrik and {O'Steen}, Richard and {Oman}, Kyle A. and {Pacifici}, Camilla and {Pascual}, Sergio and {Pascual-Granado}, J. and {Patil}, Rohit R. and {Perren}, Gabriel I. and {Pickering}, Timothy E. and {Rastogi}, Tanuj and {Roulston}, Benjamin R. and {Ryan}, Daniel F. and {Rykoff}, Eli S. and {Sabater}, Jose and {Sakurikar}, Parikshit and {Salgado}, Jes{\'u}s and {Sanghi}, Aniket and {Saunders}, Nicholas and {Savchenko}, Volodymyr and {Schwardt}, Ludwig and {Seifert-Eckert}, Michael and {Shih}, Albert Y. and {Jain}, Anany Shrey and {Shukla}, Gyanendra and {Sick}, Jonathan and {Simpson}, Chris and {Singanamalla}, Sudheesh and {Singer}, Leo P. and {Singhal}, Jaladh and {Sinha}, Manodeep and {Sip{\H{o}}cz}, Brigitta M. and {Spitler}, Lee R. and {Stansby}, David and {Streicher}, Ole and {{\v{S}}umak}, Jani and {Swinbank}, John D. and {Taranu}, Dan S. and {Tewary}, Nikita and {Tremblay}, Grant R. and {de Val-Borro}, Miguel and {Van Kooten}, Samuel J. and {Vasovi{\'c}}, Zlatan and {Verma}, Shresth and {de Miranda Cardoso}, Jos{\'e} Vin{\'\i}cius and {Williams}, Peter K.~G. and {Wilson}, Tom J. and {Winkel}, Benjamin and {Wood-Vasey}, W.~M. and {Xue}, Rui and {Yoachim}, Peter and {Zhang}, Chen and {Zonca}, Andrea and {Astropy Project Contributors}},
        title = "{The Astropy Project: Sustaining and Growing a Community-oriented Open-source Project and the Latest Major Release (v5.0) of the Core Package}",
      journal = {The Astrophysical Journal},
         year = 2022,
        month = aug,
       volume = {935},
       number = {2},
          eid = {167},
        pages = {167},
          doi = {10.3847/1538-4357/ac7c74},
archivePrefix = {arXiv},
       eprint = {2206.14220},
 primaryClass = {astro-ph.IM},
       adsurl = {https://ui.adsabs.harvard.edu/abs/2022ApJ...935..167A}
}

@article{Shi21,
author = {Shi, Chaojun and Zhou, Yatong and Qiu, Bo},
year = {2021},
month = {08},
pages = {1-14},
title = {CloudU-Netv2: A Cloud Segmentation Method for Ground-Based Cloud Images Based on Deep Learning},
volume = {53},
journal = {Neural Processing Letters},
doi = {10.1007/s11063-021-10457-2}
}

@ARTICLE{Xie20,
       author = {{Xie}, Wanyi and {Liu}, Dong and {Yang}, Ming and {Chen}, Shaoqing and {Wang}, Benge and {Wang}, Zhenzhu and {Xia}, Yingwei and {Liu}, Yong and {Wang}, Yiren and {Zhang}, Chaofang},
        title = "{SegCloud: a novel cloud image segmentation model using a deep convolutional neural network for ground-based all-sky-view camera observation}",
      journal = {Atmospheric Measurement Techniques},
         year = 2020,
        month = apr,
       volume = {13},
       number = {4},
        pages = {1953-1961},
          doi = {10.5194/amt-13-1953-2020},
       adsurl = {https://ui.adsabs.harvard.edu/abs/2020AMT....13.1953X}
}

@ARTICLE{Dev17a,
       author = {{Dev}, Soumyabrata and {Lee}, Yee Hui and {Winkler}, Stefan},
        title = "{Color-Based Segmentation of Sky/Cloud Images From Ground-Based Cameras}",
      journal = {IEEE Journal of Selected Topics in Applied Earth Observations and Remote Sensing},
         year = 2017,
        month = jan,
       volume = {10},
       number = {1},
        pages = {231-242},
          doi = {10.1109/JSTARS.2016.2558474},
archivePrefix = {arXiv},
       eprint = {1606.03669},
 primaryClass = {cs.CV},
       adsurl = {https://ui.adsabs.harvard.edu/abs/2017IJSTA..10..231D}
}

@ARTICLE{Dev17b,
       author = {{Dev}, Soumyabrata and {Savoy}, Florian M. and {Lee}, Yee Hui and {Winkler}, Stefan},
        title = "{Nighttime sky/cloud image segmentation}",
      journal = {arXiv e-prints},
         year = 2017,
        month = may,
          eid = {arXiv:1705.10583},
        pages = {arXiv:1705.10583},
          doi = {10.48550/arXiv.1705.10583},
archivePrefix = {arXiv},
       eprint = {1705.10583},
 primaryClass = {cs.CV},
       adsurl = {https://ui.adsabs.harvard.edu/abs/2017arXiv170510583D}
}

@Article{Li22,
AUTHOR = {Li, X. and Wang, B. and Qiu, B. and Wu, C.},
TITLE = {An all-sky camera image classification method using cloud cover features},
JOURNAL = {Atmospheric Measurement Techniques},
VOLUME = {15},
YEAR = {2022},
NUMBER = {11},
PAGES = {3629--3639},
URL = {https://amt.copernicus.org/articles/15/3629/2022/},
DOI = {10.5194/amt-15-3629-2022}
}

@ARTICLE{Jia25,
       author = {{Yin}, Jia and {Yao}, Yongqiang and {Qian}, Xuan and {Liu}, Liyong and {Chen}, Xu and {Zhai}, Liuming},
        title = "{Calibration and applications of the all-sky camera at the Ali Observatory in Tibet}",
      journal = {Monthly Notices of the Royal Astronomical Society},
         year = 2025,
        month = feb,
       volume = {537},
       number = {1},
        pages = {617-627},
          doi = {10.1093/mnras/staf056},
archivePrefix = {arXiv},
       eprint = {2501.08358},
 primaryClass = {astro-ph.IM},
       adsurl = {https://ui.adsabs.harvard.edu/abs/2025MNRAS.537..617Y}
}

@ARTICLE{Lang10,
       author = {{Lang}, Dustin and {Hogg}, David W. and {Mierle}, Keir and {Blanton}, Michael and {Roweis}, Sam},
        title = "{Astrometry.net: Blind Astrometric Calibration of Arbitrary Astronomical Images}",
      journal = {The Astronomical Journal},
         year = 2010,
        month = may,
       volume = {139},
       number = {5},
        pages = {1782-1800},
          doi = {10.1088/0004-6256/139/5/1782},
archivePrefix = {arXiv},
       eprint = {0910.2233},
 primaryClass = {astro-ph.IM},
       adsurl = {https://ui.adsabs.harvard.edu/abs/2010AJ....139.1782L}
}

@ARTICLE{Kannala06,
  author={Kannala, J. and Brandt, S.S.},
  journal={IEEE Transactions on Pattern Analysis and Machine Intelligence}, 
  title={A generic camera model and calibration method for conventional, wide-angle, and fish-eye lenses}, 
  year={2006},
  volume={28},
  number={8},
  pages={1335-1340},
  doi={10.1109/TPAMI.2006.153}}

@ARTICLE{Grosky05,
       author = {{G{\'o}rski}, K.~M. and {Hivon}, E. and {Banday}, A.~J. and {Wandelt}, B.~D. and {Hansen}, F.~K. and {Reinecke}, M. and {Bartelmann}, M.},
        title = "{HEALPix: A Framework for High-Resolution Discretization and Fast Analysis of Data Distributed on the Sphere}",
      journal = {The Astrophysical Journal},
         year = 2005,
        month = apr,
       volume = {622},
       number = {2},
        pages = {759-771},
          doi = {10.1086/427976},
archivePrefix = {arXiv},
       eprint = {astro-ph/0409513},
 primaryClass = {astro-ph},
       adsurl = {https://ui.adsabs.harvard.edu/abs/2005ApJ...622..759G}
}

@ARTICLE{He21,
       author = {{He}, Kaiming and {Chen}, Xinlei and {Xie}, Saining and {Li}, Yanghao and {Doll{\'a}r}, Piotr and {Girshick}, Ross},
        title = "{Masked Autoencoders Are Scalable Vision Learners}",
      journal = {arXiv e-prints},
         year = 2021,
        month = nov,
          eid = {arXiv:2111.06377},
        pages = {arXiv:2111.06377},
          doi = {10.48550/arXiv.2111.06377},
archivePrefix = {arXiv},
       eprint = {2111.06377},
 primaryClass = {cs.CV},
       adsurl = {https://ui.adsabs.harvard.edu/abs/2021arXiv211106377H}
}

@ARTICLE{Chen21,
       author = {{Chen}, Xinlei and {Xie}, Saining and {He}, Kaiming},
        title = "{An Empirical Study of Training Self-Supervised Vision Transformers}",
      journal = {arXiv e-prints},
         year = 2021,
        month = apr,
          eid = {arXiv:2104.02057},
        pages = {arXiv:2104.02057},
          doi = {10.48550/arXiv.2104.02057},
archivePrefix = {arXiv},
       eprint = {2104.02057},
 primaryClass = {cs.CV},
       adsurl = {https://ui.adsabs.harvard.edu/abs/2021arXiv210402057C}
}

@ARTICLE{Zhou21,
       author = {{Zhou}, Jinghao and {Wei}, Chen and {Wang}, Huiyu and {Shen}, Wei and {Xie}, Cihang and {Yuille}, Alan and {Kong}, Tao},
        title = "{iBOT: Image BERT Pre-Training with Online Tokenizer}",
      journal = {arXiv e-prints},
         year = 2021,
        month = nov,
          eid = {arXiv:2111.07832},
        pages = {arXiv:2111.07832},
          doi = {10.48550/arXiv.2111.07832},
archivePrefix = {arXiv},
       eprint = {2111.07832},
 primaryClass = {cs.CV},
       adsurl = {https://ui.adsabs.harvard.edu/abs/2021arXiv211107832Z}
}

@ARTICLE{Oquab23,
       author = {{Oquab}, Maxime and {Darcet}, Timoth{\'e}e and {Moutakanni}, Th{\'e}o and {Vo}, Huy and {Szafraniec}, Marc and {Khalidov}, Vasil and {Fernandez}, Pierre and {Haziza}, Daniel and {Massa}, Francisco and {El-Nouby}, Alaaeldin and {Assran}, Mahmoud and {Ballas}, Nicolas and {Galuba}, Wojciech and {Howes}, Russell and {Huang}, Po-Yao and {Li}, Shang-Wen and {Misra}, Ishan and {Rabbat}, Michael and {Sharma}, Vasu and {Synnaeve}, Gabriel and {Xu}, Hu and {Jegou}, Herv{\'e} and {Mairal}, Julien and {Labatut}, Patrick and {Joulin}, Armand and {Bojanowski}, Piotr},
        title = "{DINOv2: Learning Robust Visual Features without Supervision}",
      journal = {arXiv e-prints},
         year = 2023,
        month = apr,
          eid = {arXiv:2304.07193},
        pages = {arXiv:2304.07193},
          doi = {10.48550/arXiv.2304.07193},
archivePrefix = {arXiv},
       eprint = {2304.07193},
 primaryClass = {cs.CV},
       adsurl = {https://ui.adsabs.harvard.edu/abs/2023arXiv230407193O}
}

@ARTICLE{Oriane25,
       author = {{Sim{\'e}oni}, Oriane and {Vo}, Huy V. and {Seitzer}, Maximilian and {Baldassarre}, Federico and {Oquab}, Maxime and {Jose}, Cijo and {Khalidov}, Vasil and {Szafraniec}, Marc and {Yi}, Seungeun and {Ramamonjisoa}, Micha{\"e}l and {Massa}, Francisco and {Haziza}, Daniel and {Wehrstedt}, Luca and {Wang}, Jianyuan and {Darcet}, Timoth{\'e}e and {Moutakanni}, Th{\'e}o and {Sentana}, Leonel and {Roberts}, Claire and {Vedaldi}, Andrea and {Tolan}, Jamie and {Brandt}, John and {Couprie}, Camille and {Mairal}, Julien and {J{\'e}gou}, Herv{\'e} and {Labatut}, Patrick and {Bojanowski}, Piotr},
        title = "{DINOv3}",
      journal = {arXiv e-prints},
         year = 2025,
        month = aug,
          eid = {arXiv:2508.10104},
        pages = {arXiv:2508.10104},
          doi = {10.48550/arXiv.2508.10104},
archivePrefix = {arXiv},
       eprint = {2508.10104},
 primaryClass = {cs.CV},
       adsurl = {https://ui.adsabs.harvard.edu/abs/2025arXiv250810104S}
}

@ARTICLE{Shi25,
       author = {{Shi}, Xingjian and {Chen}, Zhourong and {Wang}, Hao and {Yeung}, Dit-Yan and {Wong}, Wai-kin and {Woo}, Wang-chun},
        title = "{Convolutional LSTM Network: A Machine Learning Approach for Precipitation Nowcasting}",
      journal = {arXiv e-prints},
         year = 2015,
        month = jun,
          eid = {arXiv:1506.04214},
        pages = {arXiv:1506.04214},
          doi = {10.48550/arXiv.1506.04214},
archivePrefix = {arXiv},
       eprint = {1506.04214},
 primaryClass = {cs.CV},
       adsurl = {https://ui.adsabs.harvard.edu/abs/2015arXiv150604214S}
}

@ARTICLE{Yan21,
       author = {{Yan}, Wilson and {Zhang}, Yunzhi and {Abbeel}, Pieter and {Srinivas}, Aravind},
        title = "{VideoGPT: Video Generation using VQ-VAE and Transformers}",
      journal = {arXiv e-prints},
         year = 2021,
        month = apr,
          eid = {arXiv:2104.10157},
        pages = {arXiv:2104.10157},
          doi = {10.48550/arXiv.2104.10157},
archivePrefix = {arXiv},
       eprint = {2104.10157},
 primaryClass = {cs.CV},
       adsurl = {https://ui.adsabs.harvard.edu/abs/2021arXiv210410157Y}
}

@article{Hamill93,
author = {Hamill, Thomas and Nehrkorn, T.},
year = {1993},
month = {12},
pages = {401-411},
title = {A Short-Term Cloud Forecast Scheme Using Cross Correlations},
volume = {8},
journal = {Weather and Forecasting - WEATHER FORECAST},
doi = {10.1175/1520-0434(1993)008<0401:ASTCFS>2.0.CO;2}
}

@InProceedings{Farneb03,
author="Farneb{\"a}ck, Gunnar",
editor="Bigun, Josef
and Gustavsson, Tomas",
title="Two-Frame Motion Estimation Based on Polynomial Expansion",
booktitle="Image Analysis",
year="2003",
publisher="Springer Berlin Heidelberg",
address="Berlin, Heidelberg",
pages="363--370",
isbn="978-3-540-45103-7"
}

@ARTICLE{Rui25,
       author = {{Rui}, Yicheng and {Xuan}, Yifan and {Zheng}, Shuyue and {Li}, Kexin and {Cui}, Kaiming and {Xiao}, Kai and {Zheng}, Jie and {Ng}, Jun Kai and {Jiang}, Hongxuan and {Feng}, Fabo and {Sun}, Qinghui},
        title = "{Architecture of the Tianyu Software: Relative Photometry as a Case Study}",
      journal = {Publications of the Astronomical Society of the Pacific},
         year = 2025,
        month = jun,
       volume = {137},
       number = {6},
          eid = {064501},
        pages = {064501},
          doi = {10.1088/1538-3873/add6d9},
archivePrefix = {arXiv},
       eprint = {2505.09107},
 primaryClass = {astro-ph.IM},
       adsurl = {https://ui.adsabs.harvard.edu/abs/2025PASP..137f4501R}
}

@software{Kentaro21,
  author       = {Kentaro Wada and
                  mpitid and
                  Martijn Buijs and
                  Zhang Ch. N. and
                  Narumi and
                  Bc. Martin Kubovčík and
                  Alex Myczko and
                  latentix and
                  Lingjie Zhu and
                  Naoya Yamaguchi and
                  Shohei Fujii and
                  iamgd67 and
                  IlyaOvodov and
                  Akshar Patel and
                  Christian Clauss and
                  Eisoku Kuroiwa and
                  Roger Iyengar and
                  Sergei Shilin and
                  Tanya Malygina and
                  Kento Kawaharazuka and
                  Jonne Engelberts and
                  Aleksi J and
                  AlexMa and
                  Changwoo Song and
                  Charlie and
                  Daniel Rose and
                  Douglas Livingstone and
                  Doug and
                  Erik and
                  Henrik Toft},
  title        = {wkentaro/labelme: v4.6.0},
  month        = nov,
  year         = 2021,
  publisher    = {Zenodo},
  version      = {v4.6.0},
  doi          = {10.5281/zenodo.5711226},
  url          = {https://doi.org/10.5281/zenodo.5711226},
}

@Article{ schmidt2025,
author = "Schmidt, Thomas and St{\"u}hrenberg, Jonas and Blum, Niklas and Lezaca, Jorge and Hammer, Annette and Wilbert, Stefan and Nouri, Bijan and Schroedter-Homscheidt, Marion and Heinemann, Detlev and Vogt, Thomas",
journal = "Meteorologische Zeitschrift",
month = 07,
year = 2025,
title = "Eye2Sky - a network of all-sky imager and meteorological measurement stations for high resolution nowcasting of solar irradiance",
number = "1",
volume = "34",
pages = {35-55},
url = "http://dx.doi.org/10.1127/metz/2025/1245",
doi = "10.1127/metz/2025/1245",
publisher = "Schweizerbart Science Publishers",
address = "Stuttgart, Germany"
}

@ARTICLE{Wang24,
       author = {{Wang}, Yuxuan and {Wu}, Haixu and {Dong}, Jiaxiang and {Liu}, Yong and {Wang}, Chen and {Long}, Mingsheng and {Wang}, Jianmin},
        title = "{Deep Time Series Models: A Comprehensive Survey and Benchmark}",
      journal = {arXiv e-prints},
         year = 2024,
        month = jul,
          eid = {arXiv:2407.13278},
        pages = {arXiv:2407.13278},
          doi = {10.48550/arXiv.2407.13278},
archivePrefix = {arXiv},
       eprint = {2407.13278},
 primaryClass = {stat.ML},
       adsurl = {https://ui.adsabs.harvard.edu/abs/2024arXiv240713278W}
}

@ARTICLE{Zhang25,
       author = {{Zhang}, Yuanzhao and {Gilpin}, William},
        title = "{Context parroting: A simple but tough-to-beat baseline for foundation models in scientific machine learning}",
      journal = {arXiv e-prints},
         year = 2025,
        month = may,
          eid = {arXiv:2505.11349},
        pages = {arXiv:2505.11349},
          doi = {10.48550/arXiv.2505.11349},
archivePrefix = {arXiv},
       eprint = {2505.11349},
 primaryClass = {stat.ML},
       adsurl = {https://ui.adsabs.harvard.edu/abs/2025arXiv250511349Z}
}

@ARTICLE{Fu24,
       author = {{Fu}, Yunxiang and {Lou}, Meng and {Yu}, Yizhou},
        title = "{SegMAN: Omni-scale Context Modeling with State Space Models and Local Attention for Semantic Segmentation}",
      journal = {arXiv e-prints},
         year = 2024,
        month = dec,
          eid = {arXiv:2412.11890},
        pages = {arXiv:2412.11890},
          doi = {10.48550/arXiv.2412.11890},
archivePrefix = {arXiv},
       eprint = {2412.11890},
 primaryClass = {cs.CV},
       adsurl = {https://ui.adsabs.harvard.edu/abs/2024arXiv241211890F}
}
}

\clearpage

\onecolumn
\setcounter{page}{1}
\maketitlesupplementary
\appendix

\section{Failure cases of the all-sky camera}
\label{sec:failurecase}
\begin{figure*}
    \centering
    \includegraphics[width = 0.9\linewidth]{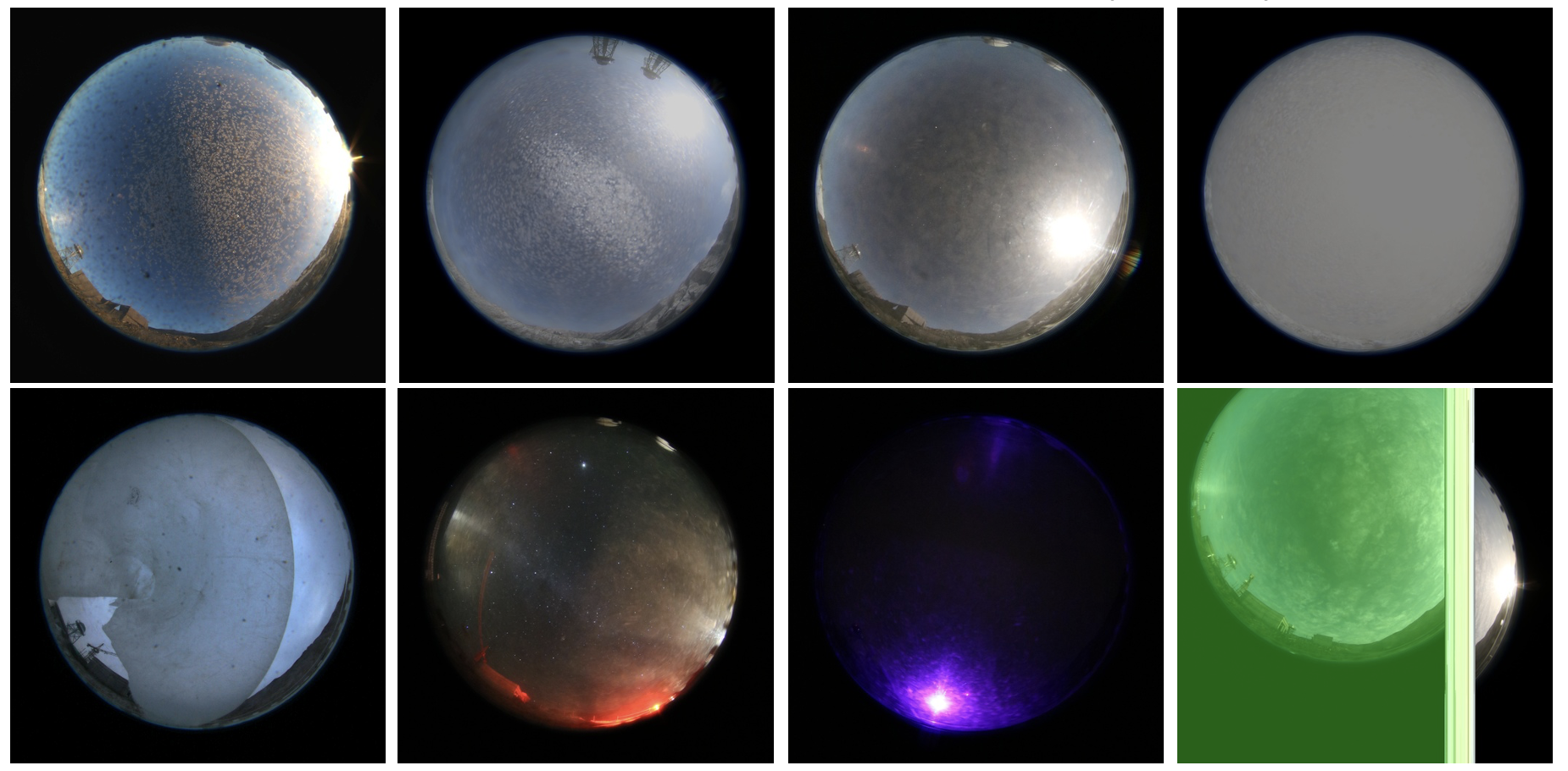}
    \caption{Failure cases of the all-sky camera.
Top row (left to right): (1) Covered by dust or sand; (2) Covered by dew or ice; (3) Scattered light caused by mud coverage; (4) Covered by snow.
Bottom row (left to right): (5) Obstruction by an external object; (6) Strong nearby light source; (7) Strong distant light source; (8) Camera malfunction.\label{fig:camerafailure}}
\end{figure*}

Failure of all-sky camera refers to being unable to recognize a large proportion of the sky condition manually. Typical cases of failure is shown in Fig.\ref{fig:camerafailure}. We manually inspect the whole dataset  and mark out the failure cases. Case 1$\sim$4 in Fig.\ref{fig:camerafailure} are marked as cover; Case 5 are marked as object; Case 6 and 7 are marked as strong light; Case 8 is marked as Camera malfunction. $665$ failures are recognized by human in the dataset. The start and the end of these events are also recorded. In segmentation task, the regions that cannot decide whether there is cloud or not are annotated as ``contamination" class. An example of ``cover" is the third column of Fig. \ref{fig:overview}. A statistics of the failure time is shown in Table~\ref{tab:failure}. Mud and dew are the most frequent causes of failure, particularly from November through May. The complete table is available in the online repository. More detailed weather monitoring and camera failure information is available in \url{https://huggingface.co/datasets/ruiyicheng/LenghuSky-8/tree/main/data}. The region that would affect the determination of the local weather are annotated as contamination, e.g. the region that is covered by dew.

\begin{table}[]
    \caption{Statistics of failure case of all-sky camera. The events that occurs within 12 hours are merged together as one event. }
    \label{tab:failure}
    \centering
\begin{tabular}{llll}
\hline
Reason & \# occur & \# frame & Down time [days] \\
\hline
cover & 181 & 30111 & 169.79 \\
stronglight & 388 & 5275 & 23.54 \\
cameradown & 56 & 2167 & 20.75 \\
object & 40 & 735 & 4.66 \\
\hline
Total & 665 & 38288 & 218.75 \\
\hline
\end{tabular}

\end{table}

\section{Projection type of all-sky camera}
In this work, several projection types are used  for distortion correction. These types are listed in table \ref{tab:full_sky_proj_type}.
\begin{table}[]
    \centering
    \caption{Optional projection types of full-sky camera, which is provided by \cite{Kannala06}.}\label{tab:full_sky_proj_type}
    \begin{tabular}{ll}
    \hline
      Name& Formula  \\\hline
       Gnomonic projection& $r=f\tan \theta$\\
      Stereographic projection& $r=2f\tan\frac{\theta}{2}$ \\
      Equidistant projection& $r=2f\theta$ \\
      Equisolid angle projection& $r=2f\sin\frac{\theta}{2}$ \\
      Orthographic projection &$r=f\sin\theta$\\
       \hline
    \end{tabular}
\end{table}

\section{Manual annotation samples and annotation process details for cloud segmentation task}
\begin{figure*}
    \centering
    \includegraphics[width=0.8\linewidth]{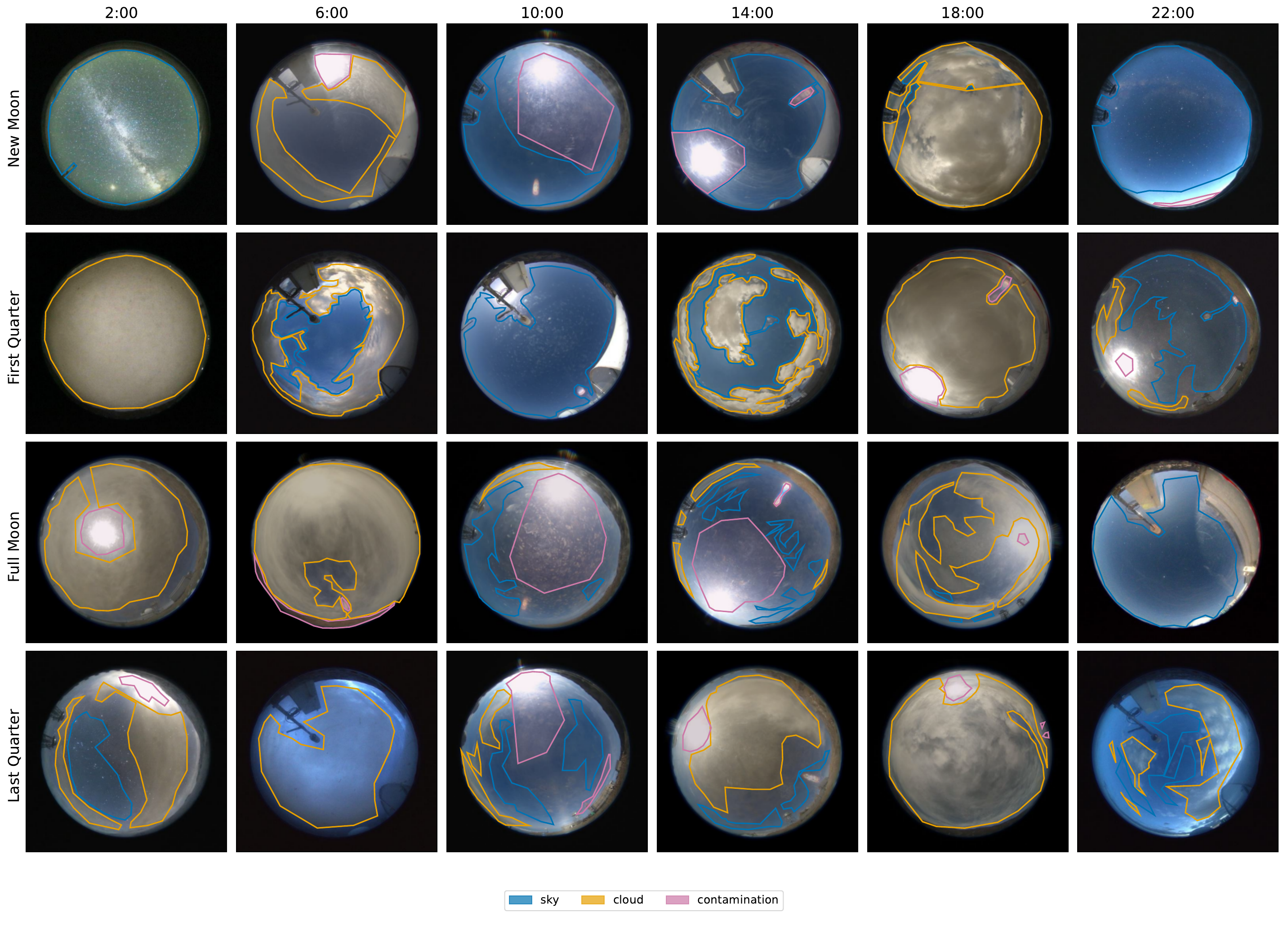}
    \caption{Manual annotation for cloud segmentation. Blue represents cloud regions; orange represents sky regions; Pink represents contamination regions; Columns represents images taken around given time in UTC+8; Rows represents images taken in different moon phase condition. Nearby frames are used by human to determine whether some regions are scatter light or cloud. }
    \label{fig:annotation_segmentation}
\end{figure*}

Manual annotation samples for cloud segmentation task is shown in Fig.~\ref{fig:annotation_segmentation}. The 1,111-image reference set was annotated by 9 trained students/engineers with astronomy background under an astronomer lead. We stratified sampling by time-of-day, season, moon phase, and cloud coverage to ensure balanced coverage; rare cases largely determine the final size. Each image is labeled in LabelMe using written guidelines with conservative partial labeling: only high-confidence regions are labeled; ambiguous pixels are left unlabeled and ignored in loss/metrics. Every image then undergoes a second-pass expert review/edit to enforce cross-annotator consistency; disagreements are resolved by adjudication in this pass.

Some bright structures in examples Fig.~\ref{fig:annotation_segmentation} arise from scattering/stray light (dust/dew/scratches/saturation) and can resemble thin clouds in a single frame. Our guideline treats these as contamination when they obstruct sky/cloud attribution; otherwise they are annotated as sky/cloud, or left unlabeled if ambiguous. Annotators also consult adjacent frames to distinguish evolving clouds from static artifacts.

\section{Fitting and residual of astrometric calibration}

\begin{figure*}[t]
  \centering

  \begin{subfigure}[t]{0.24\linewidth}
    \centering
    \includegraphics[width=\linewidth]{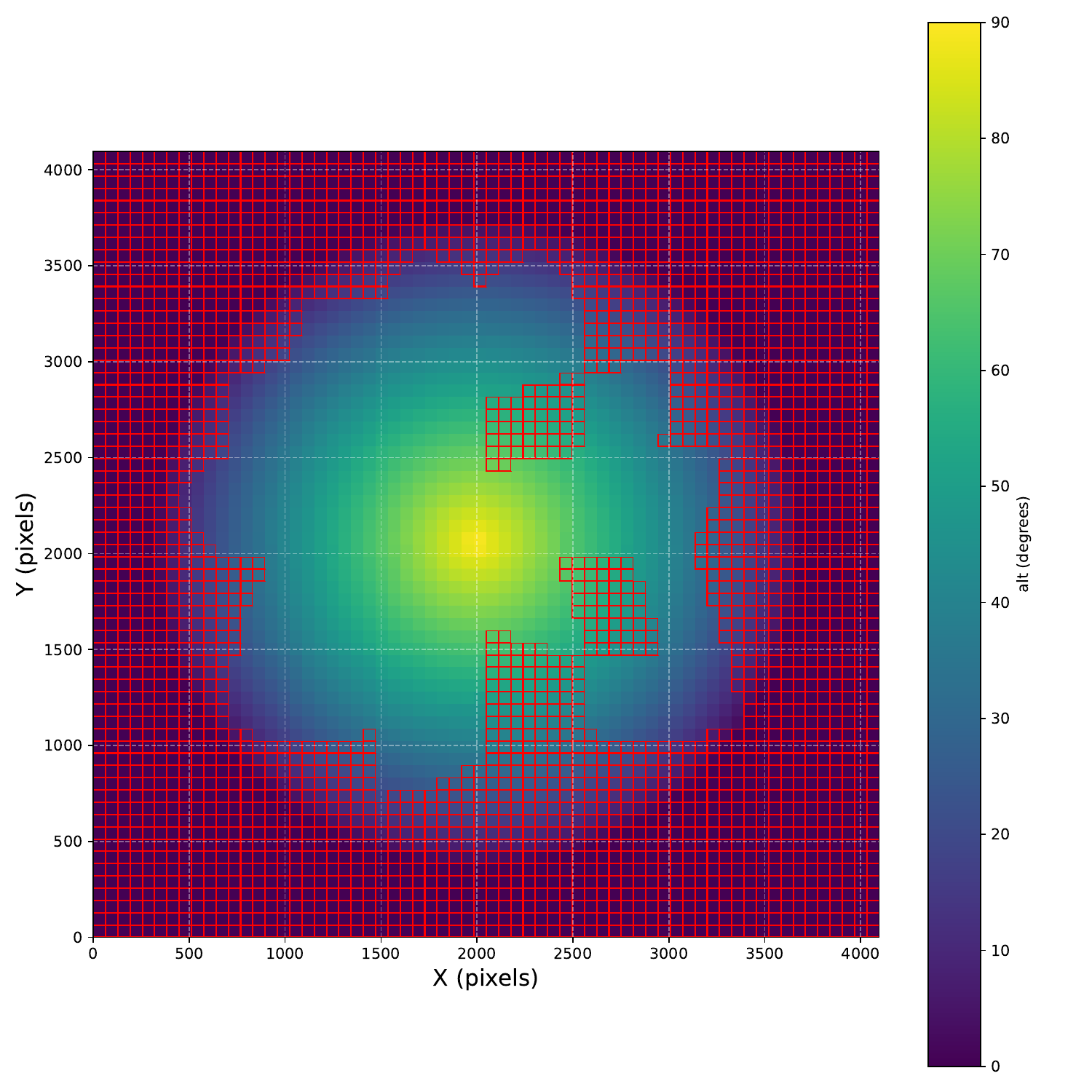}
    \caption{Altitude fitting result for 2018-05-01 00:02:44}
  \end{subfigure}\hfill
  \begin{subfigure}[t]{0.24\linewidth}
    \centering
    \includegraphics[width=\linewidth]{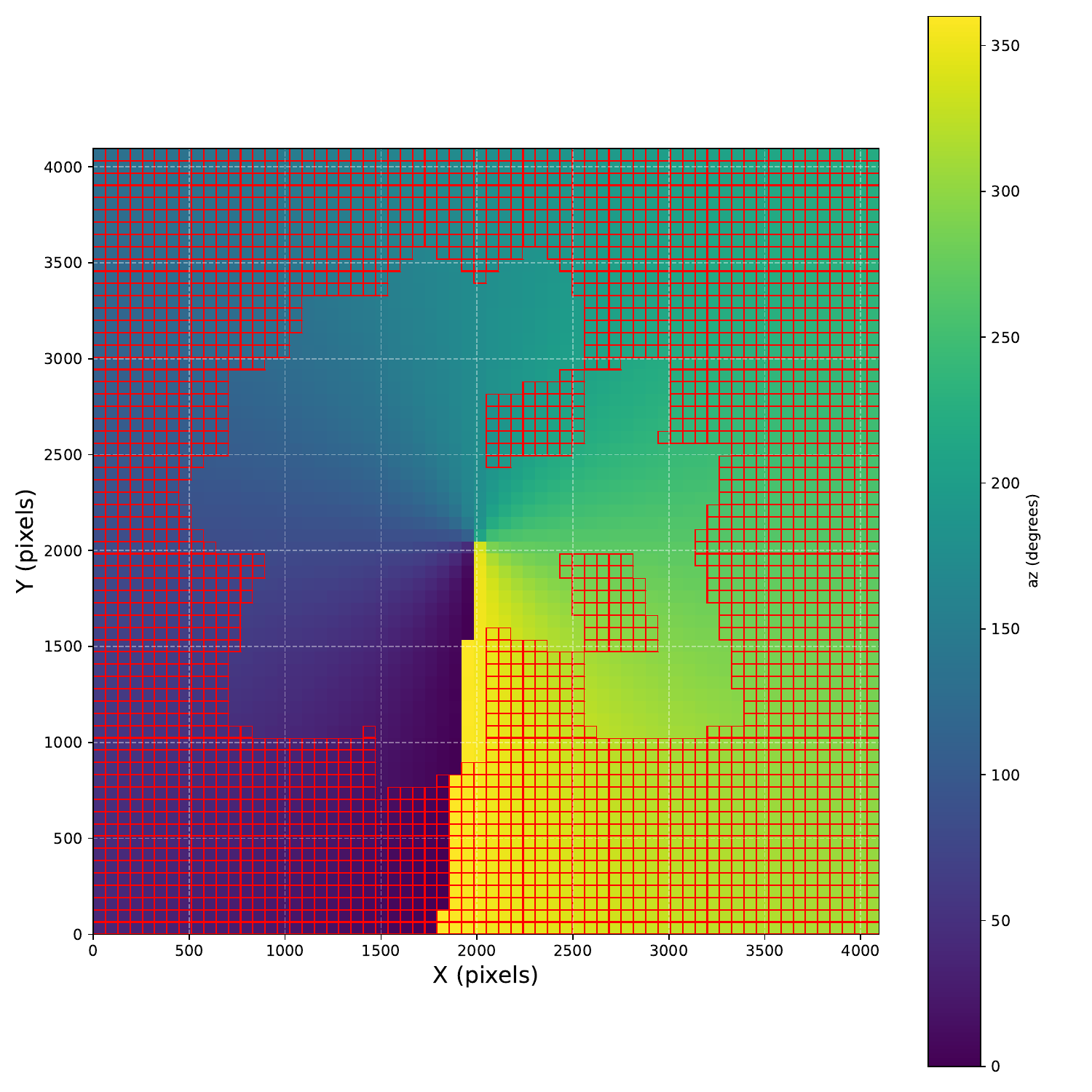}
    \caption{Azimuth fitting result for 2018-05-01 00:02:44}
  \end{subfigure}\hfill
  \begin{subfigure}[t]{0.24\linewidth}
    \centering
    \includegraphics[width=\linewidth]{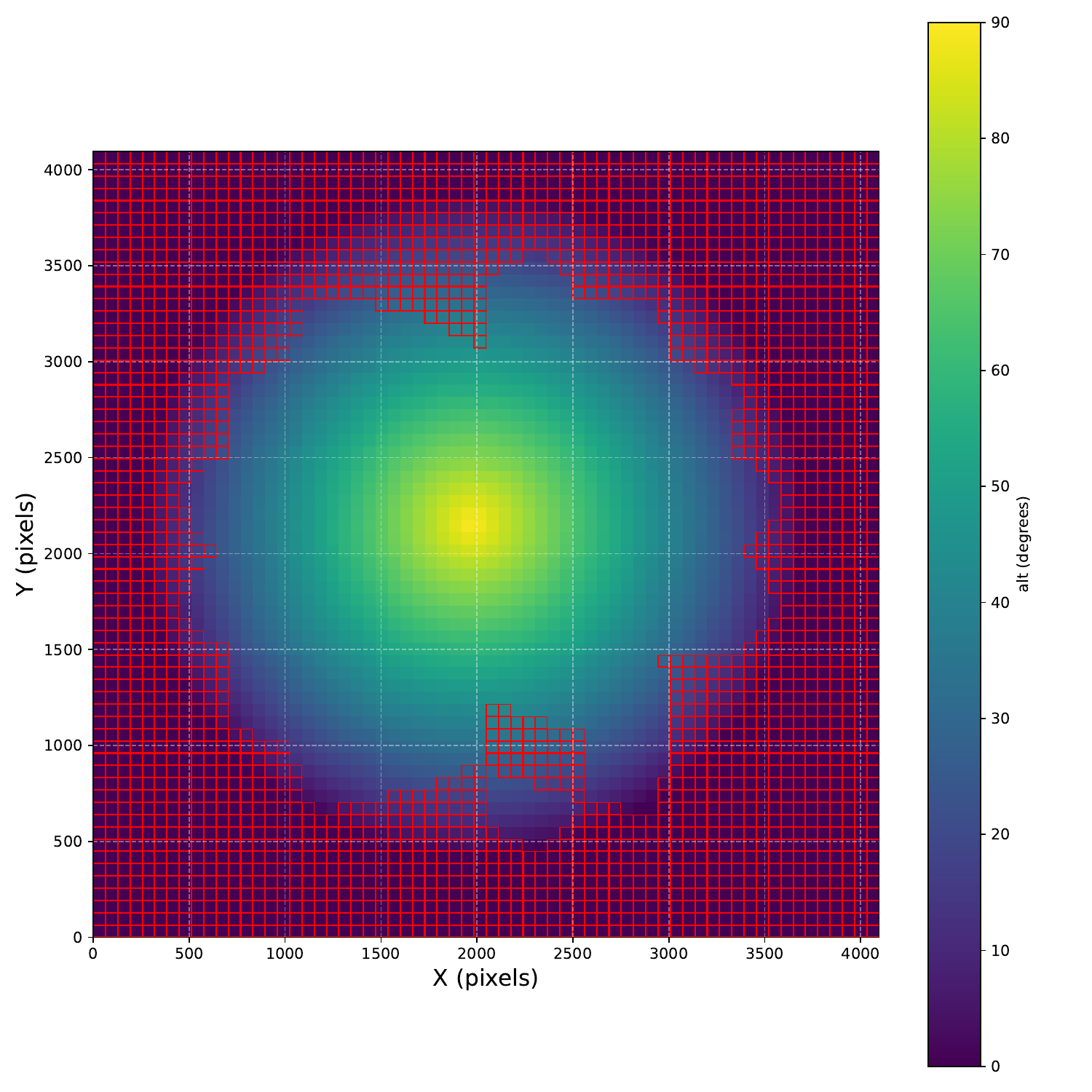}
    \caption{Altitude fitting result for 2018-09-27 19:19:49}
  \end{subfigure}\hfill
  \begin{subfigure}[t]{0.24\linewidth}
    \centering
    \includegraphics[width=\linewidth]{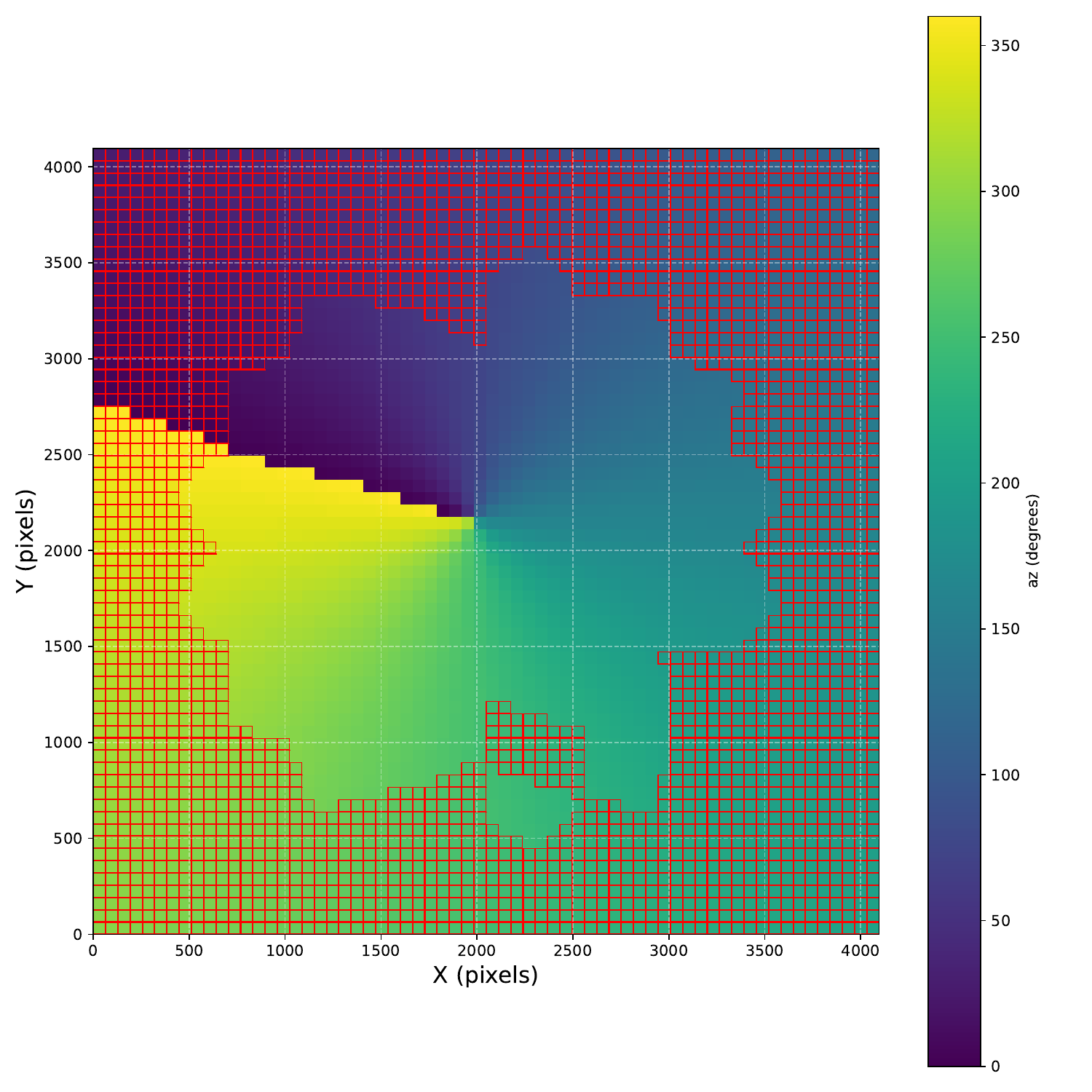}
    \caption{Azimuth fitting result for 2018-09-27 19:19:49}
  \end{subfigure}

  \vspace{0.6em}

  \begin{subfigure}[t]{0.24\linewidth}
    \centering
    \includegraphics[width=\linewidth]{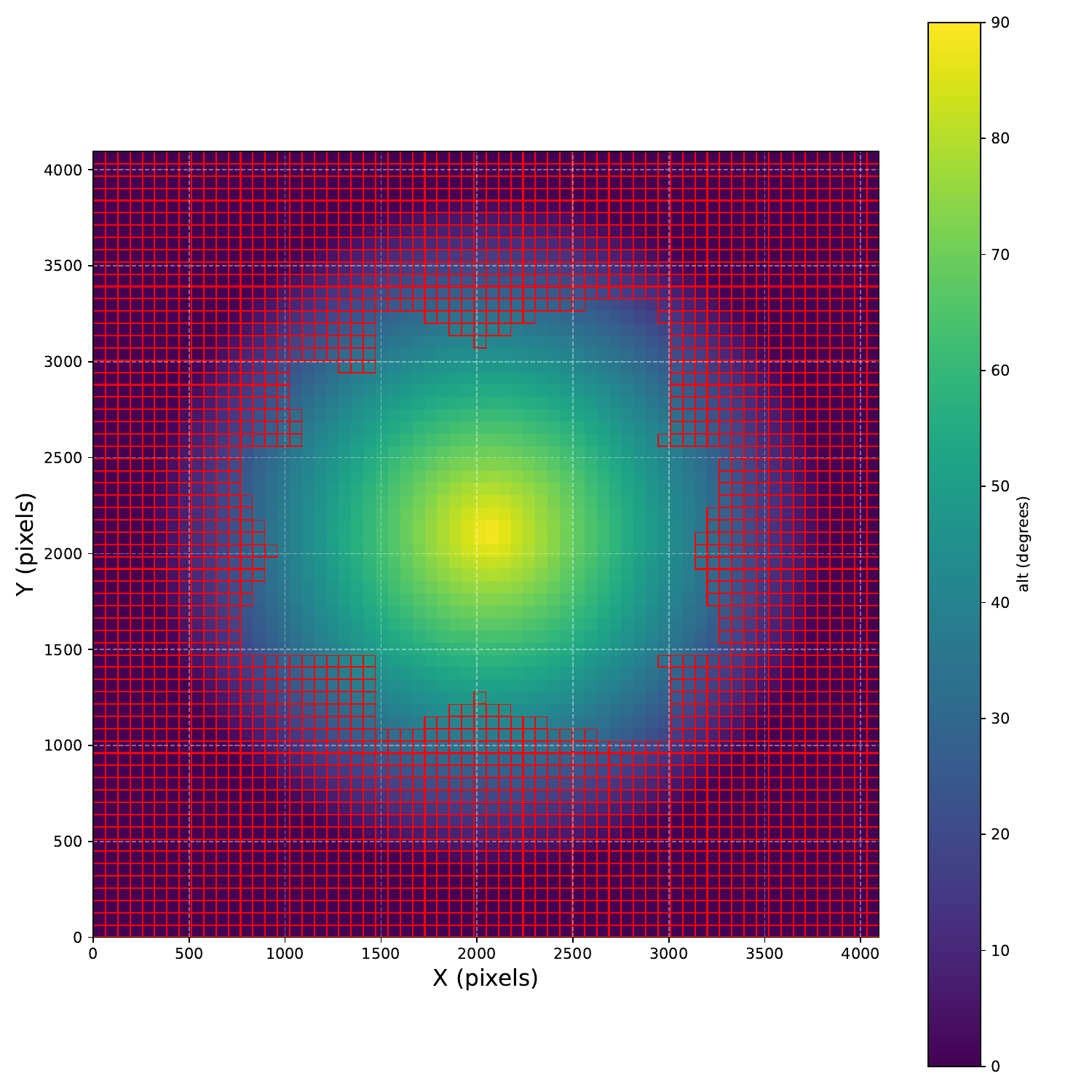}
    \caption{Altitude fitting result for 2019-04-24 15:39:36}
  \end{subfigure}\hfill
  \begin{subfigure}[t]{0.24\linewidth}
    \centering
    \includegraphics[width=\linewidth]{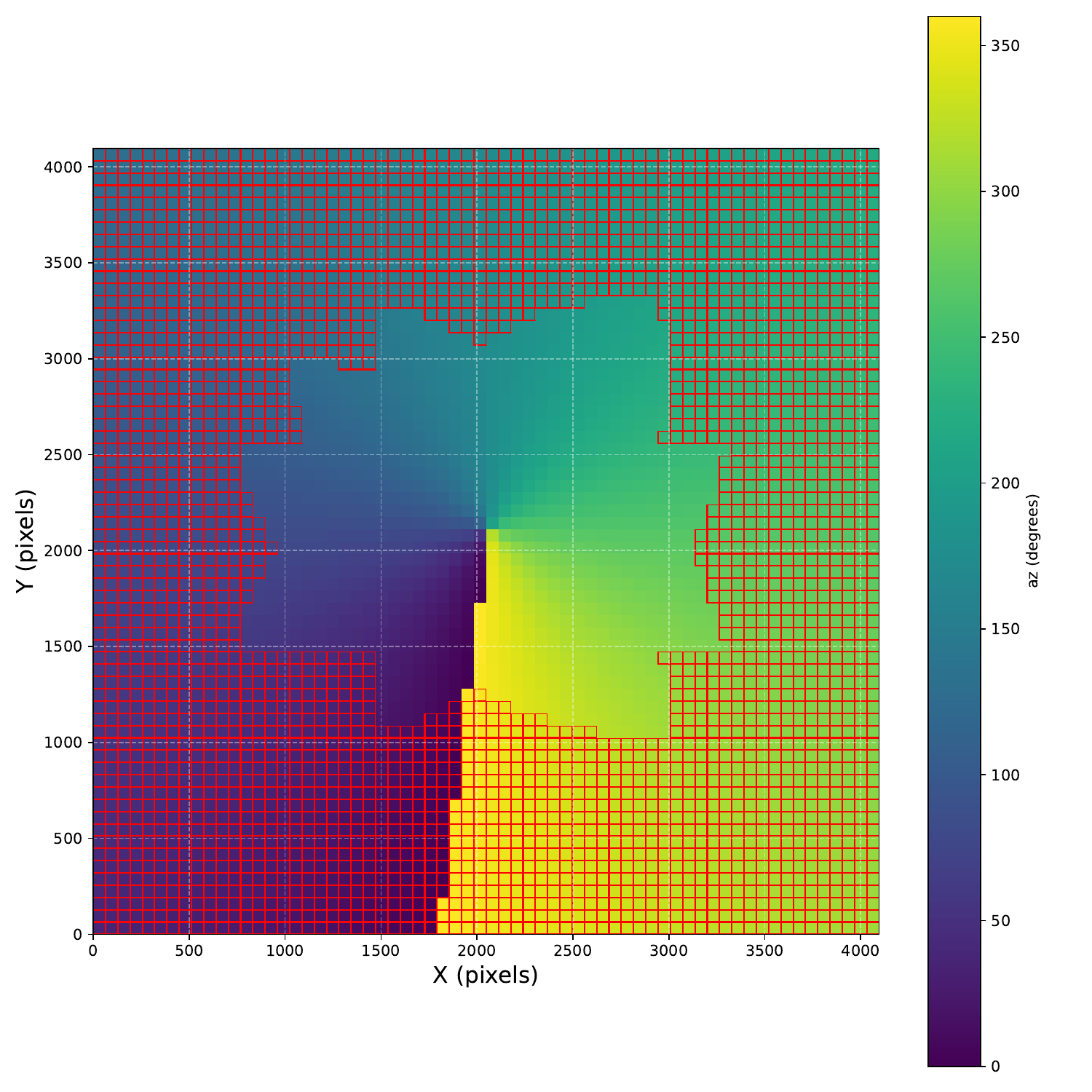}
    \caption{Azimuth fitting result for 2019-04-24 15:39:36}
  \end{subfigure}\hfill
  \begin{subfigure}[t]{0.24\linewidth}
    \centering
    \includegraphics[width=\linewidth]{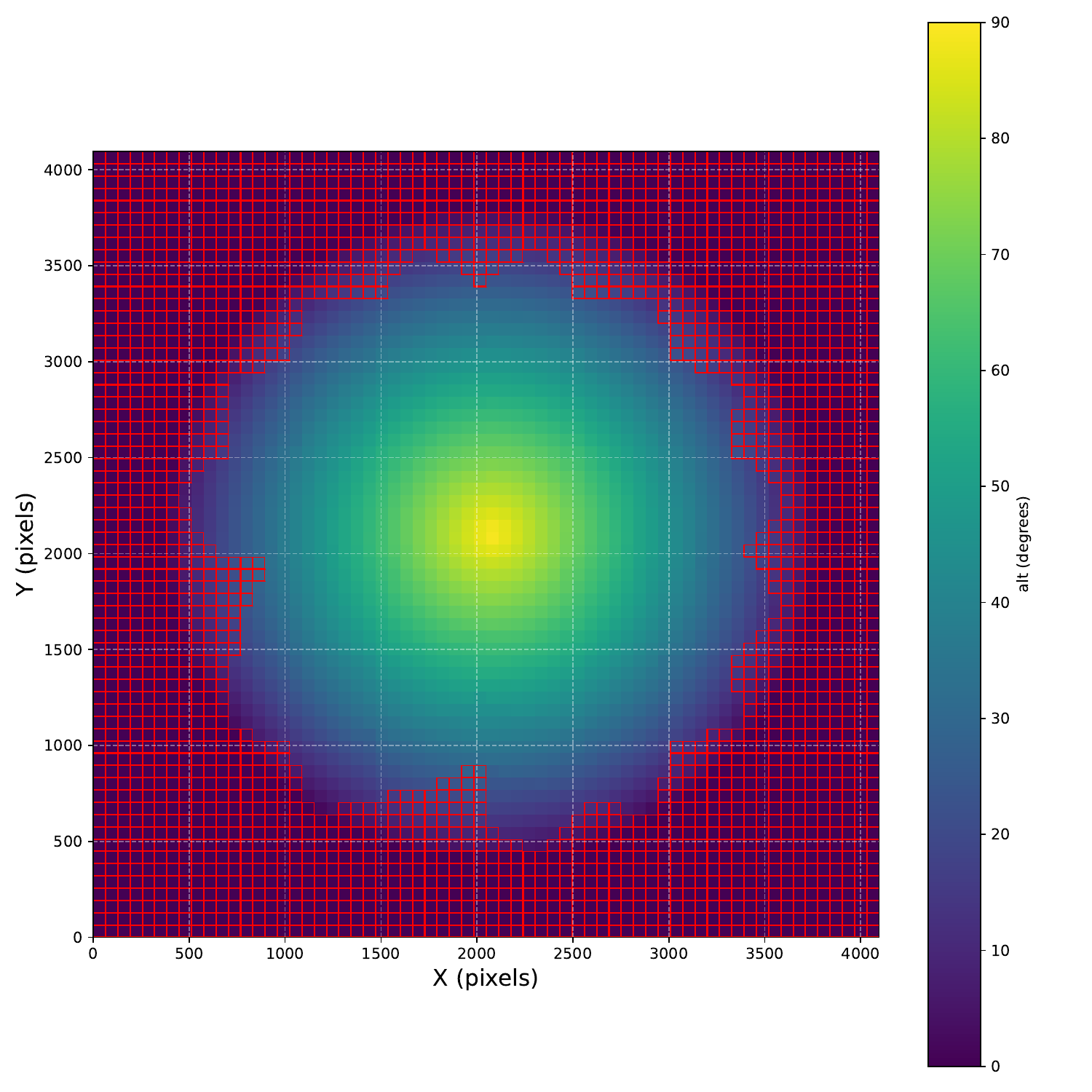}
    \caption{Altitude fitting result for 2019-06-26 18:23:18}
  \end{subfigure}\hfill
  \begin{subfigure}[t]{0.24\linewidth}
    \centering
    \includegraphics[width=\linewidth]{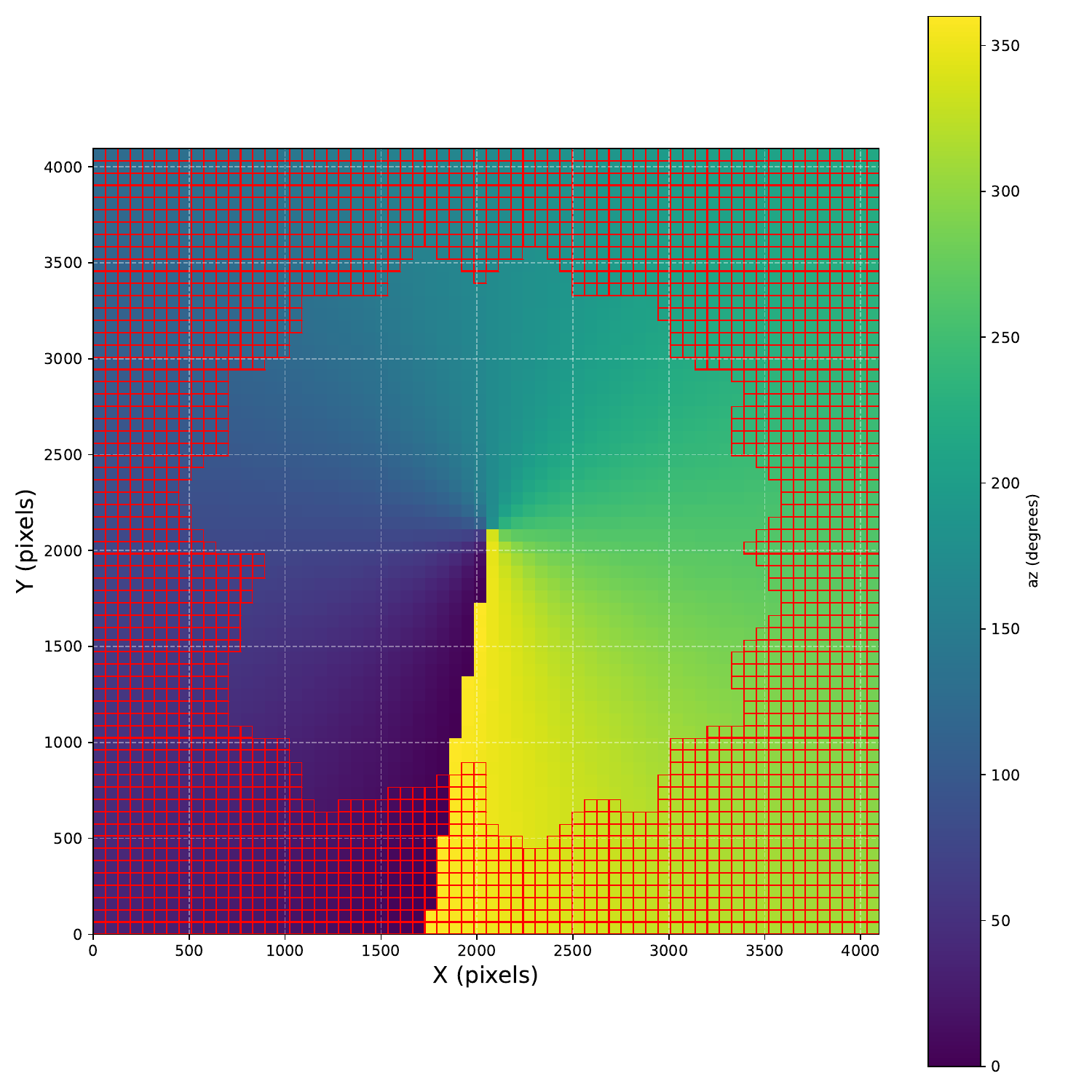}
    \caption{Azimuth fitting result for 2019-06-26 18:23:18}
  \end{subfigure}

  \vspace{0.6em}

  \begin{subfigure}[t]{0.24\linewidth}
    \centering
    \includegraphics[width=\linewidth]{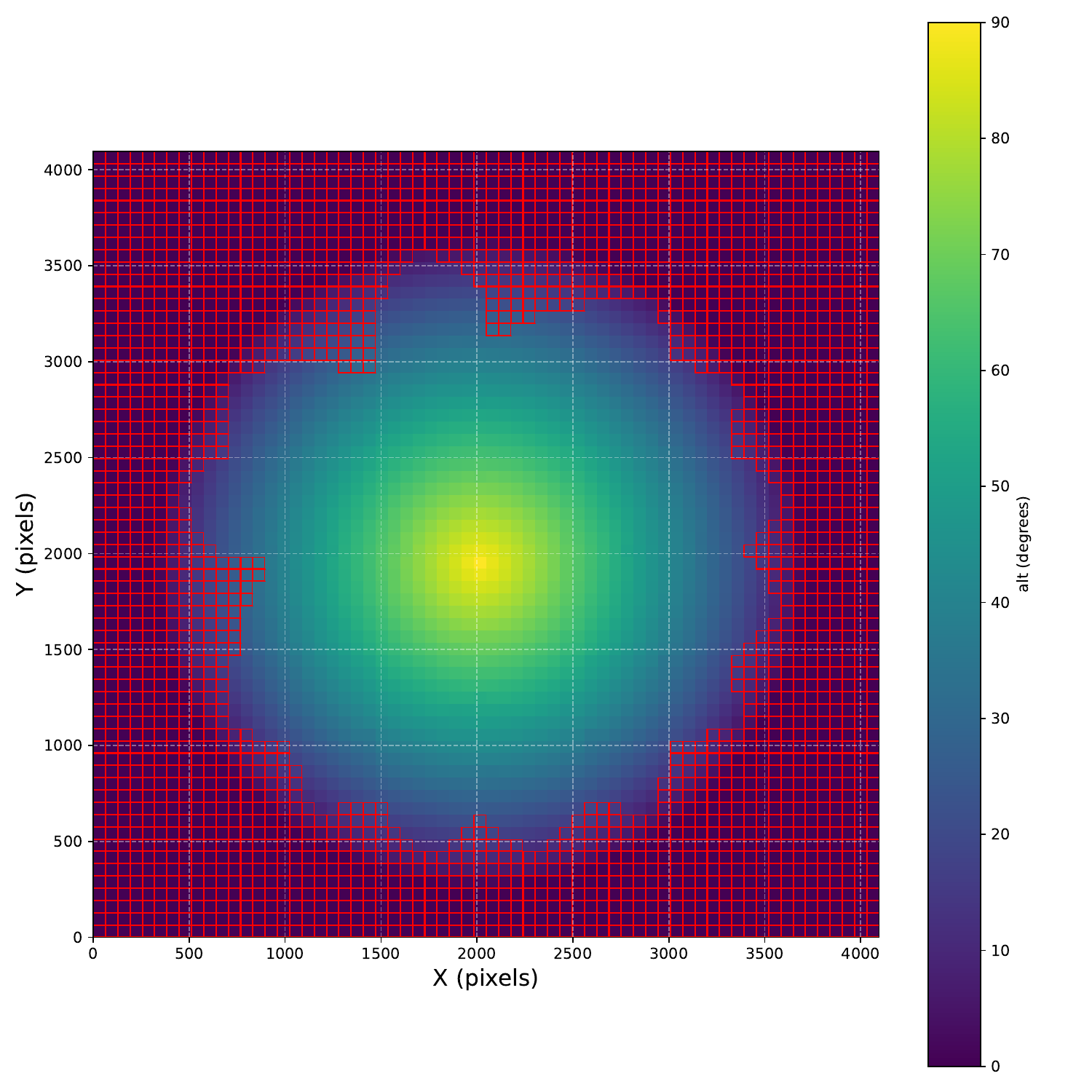}
    \caption{Altitude fitting result for 2019-07-05 11:59:14}
  \end{subfigure}\hfill
  \begin{subfigure}[t]{0.24\linewidth}
    \centering
    \includegraphics[width=\linewidth]{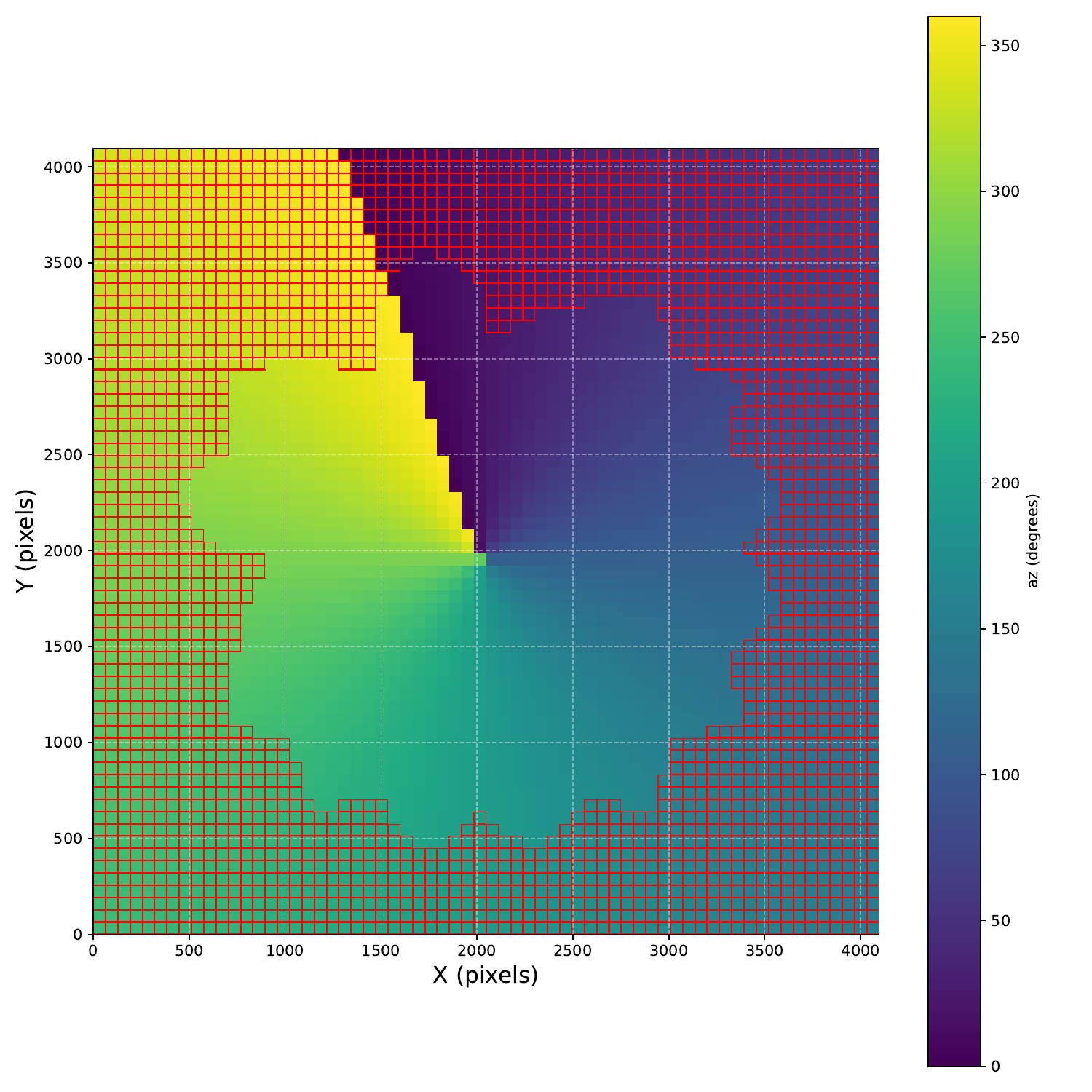}
    \caption{Azimuth fitting result for 2019-07-05 11:59:14}
  \end{subfigure}\hfill
  \begin{subfigure}[t]{0.24\linewidth}
    \centering
    \includegraphics[width=\linewidth]{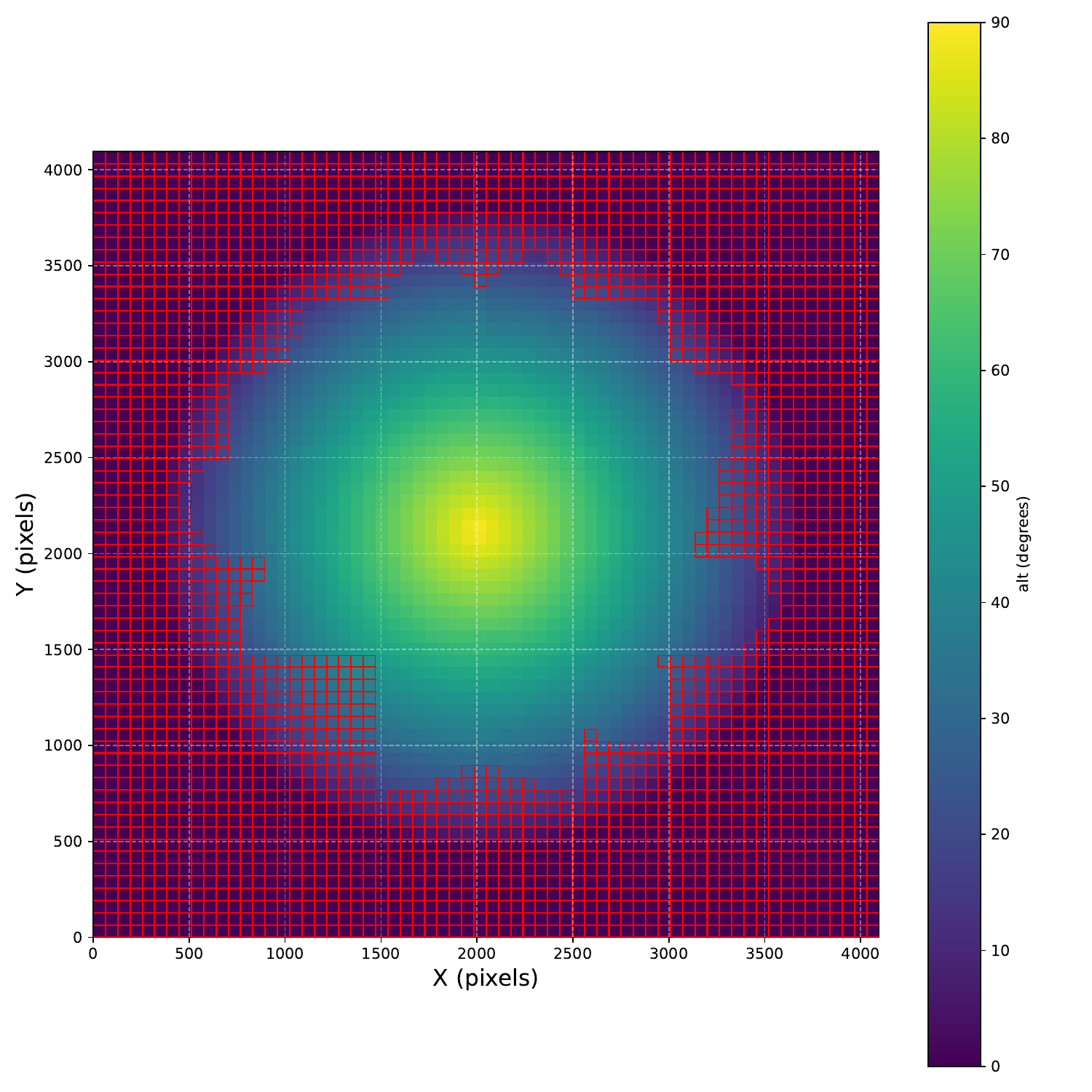}
    \caption{Altitude fitting result for 2023-09-27 18:09:48}
  \end{subfigure}\hfill
  \begin{subfigure}[t]{0.24\linewidth}
    \centering
    \includegraphics[width=\linewidth]{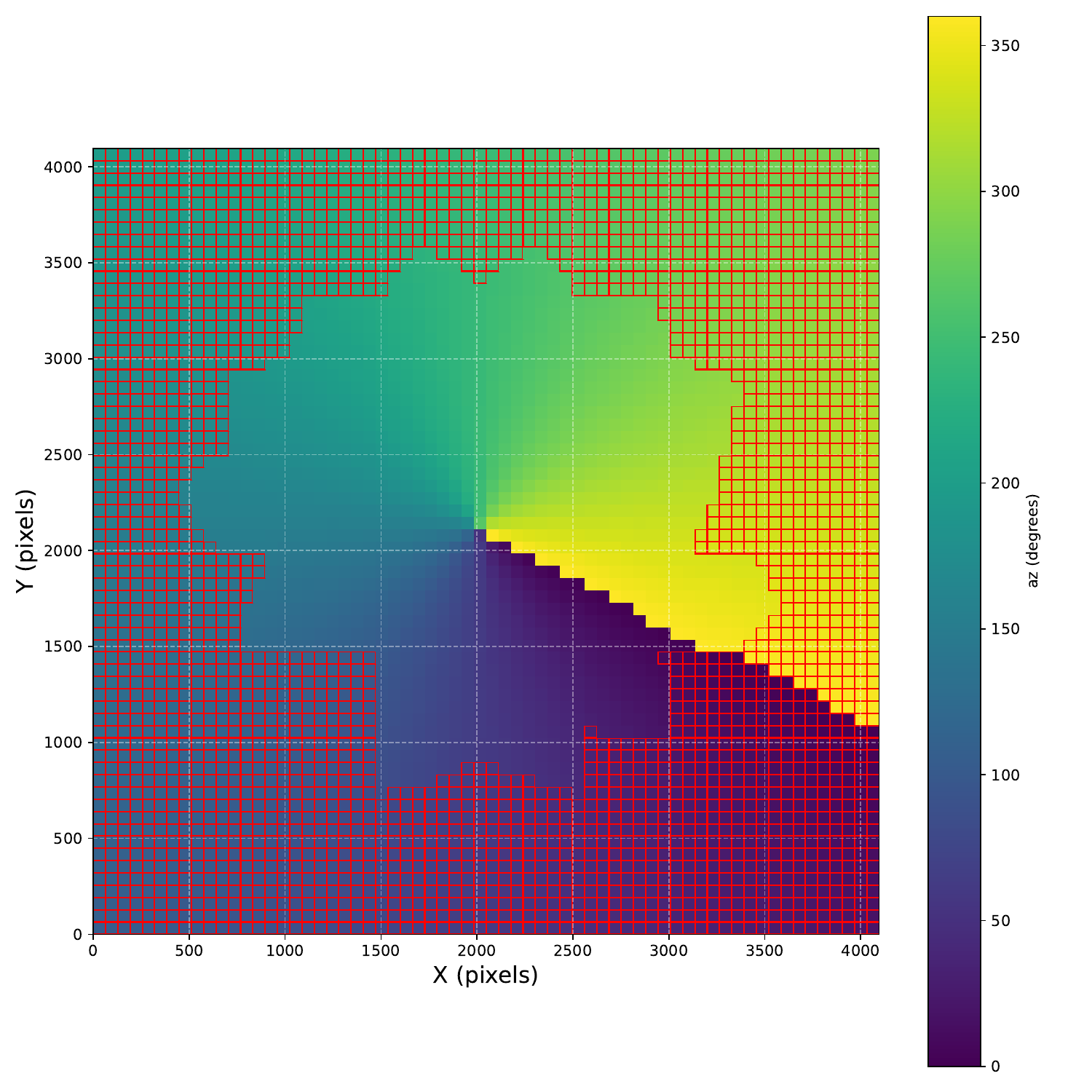}
    \caption{Azimuth fitting result for 2023-09-27 18:09:48}
  \end{subfigure}

  \vspace{0.6em}
  \caption{Altitude–Azimuth fitting results for different time slots. Red boxes means that the WCS of the corresponding HEALPix cell is not resolved by \textsc{Astrometry.net} for any image in the ensemble, and the corresponding altitude and azimuth is obtained by fitting results.}
  \label{fig:altazfit}
\end{figure*}

\begin{figure*}[t]
  \centering

  \begin{subfigure}[t]{0.24\linewidth}
    \centering
    \includegraphics[width=\linewidth]{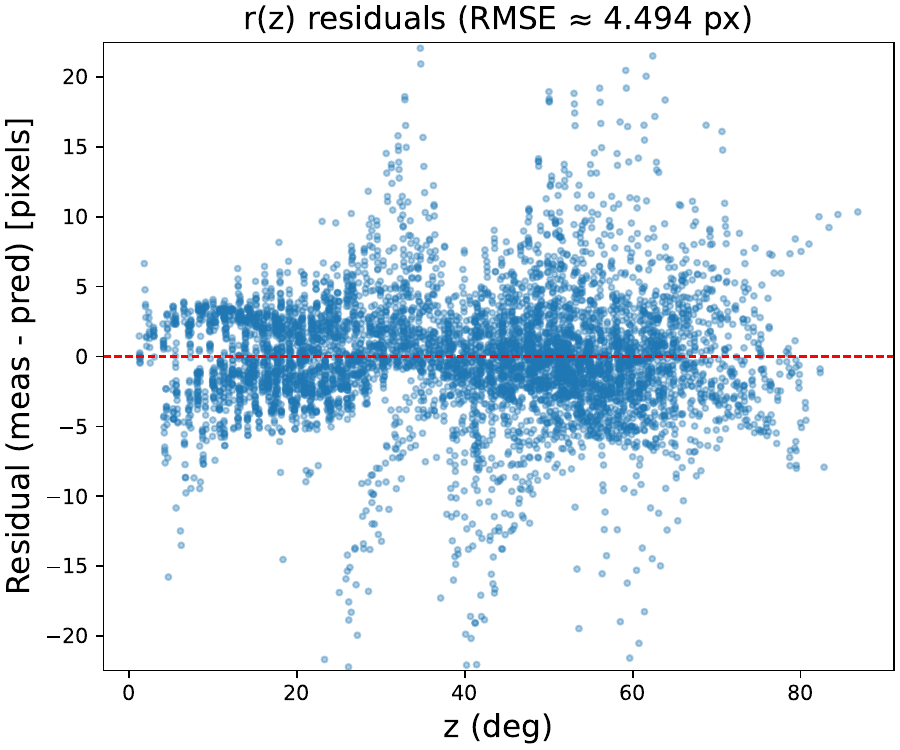}
    \caption{Radial fitting residual for 2018-05-01 00:02:44}
  \end{subfigure}\hfill
  \begin{subfigure}[t]{0.24\linewidth}
    \centering
    \includegraphics[width=\linewidth]{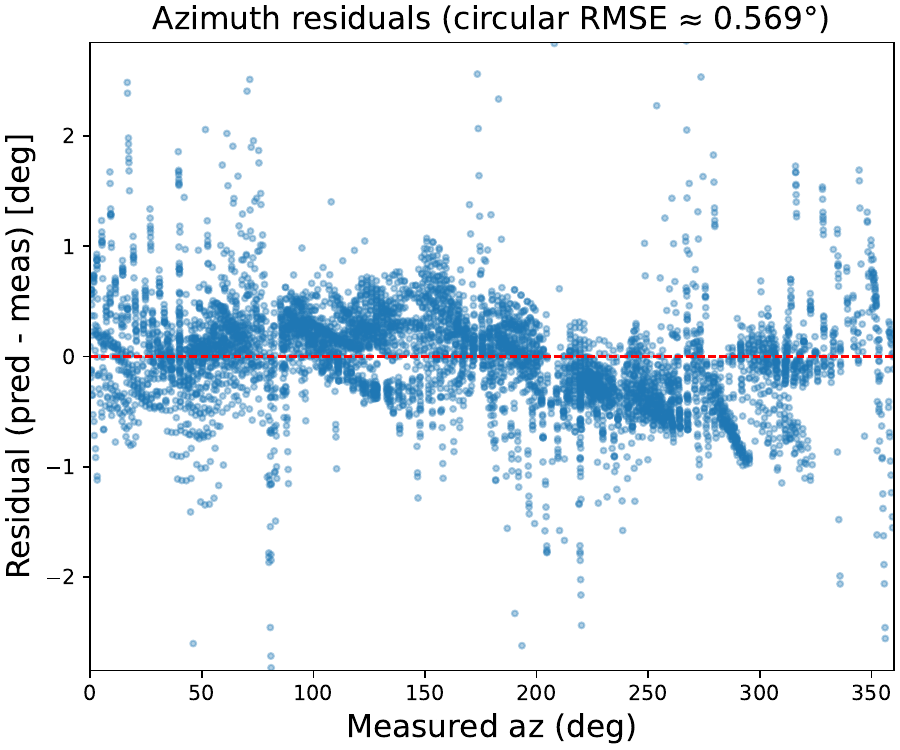}
    \caption{Azimuth fitting residual for 2018-05-01 00:02:44}
  \end{subfigure}\hfill
  \begin{subfigure}[t]{0.24\linewidth}
    \centering
    \includegraphics[width=\linewidth]{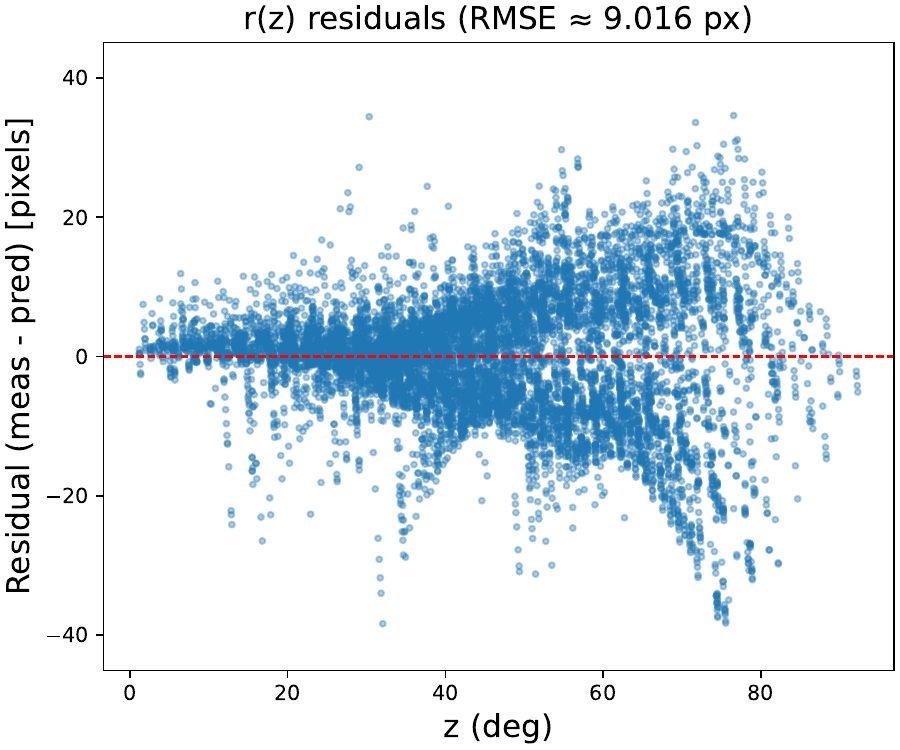}
    \caption{Radial fitting residual for 2018-09-27 19:19:49}
  \end{subfigure}\hfill
  \begin{subfigure}[t]{0.24\linewidth}
    \centering
    \includegraphics[width=\linewidth]{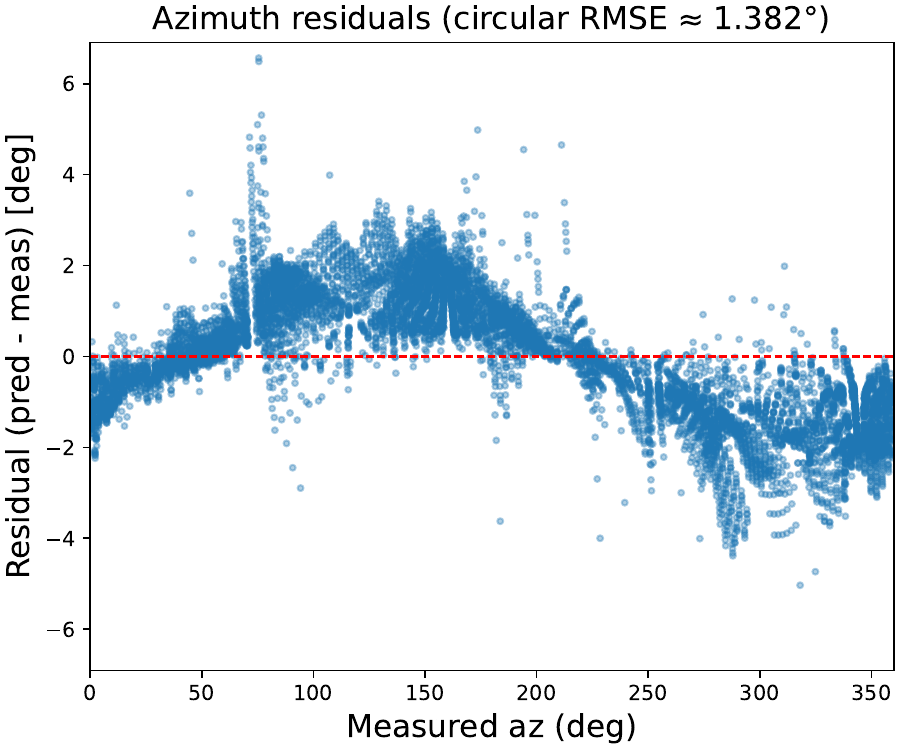}
    \caption{Azimuth fitting residual for 2018-09-27 19:19:49}
  \end{subfigure}

  \vspace{0.6em}

  \begin{subfigure}[t]{0.24\linewidth}
    \centering
    \includegraphics[width=\linewidth]{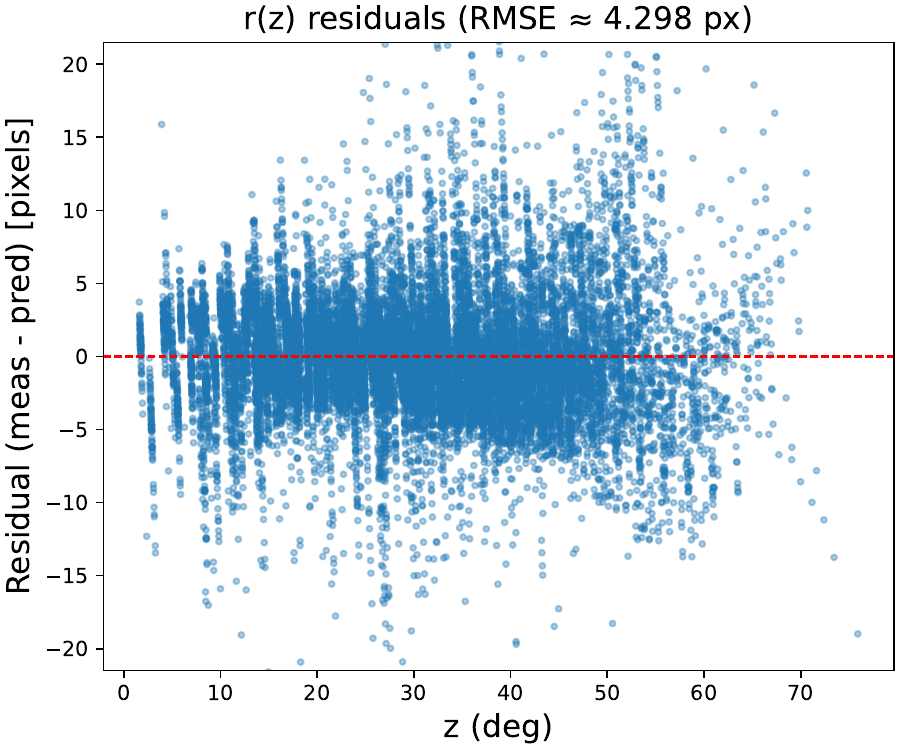}
    \caption{Radial fitting residual for 2019-04-24 15:39:36}
  \end{subfigure}\hfill
  \begin{subfigure}[t]{0.24\linewidth}
    \centering
    \includegraphics[width=\linewidth]{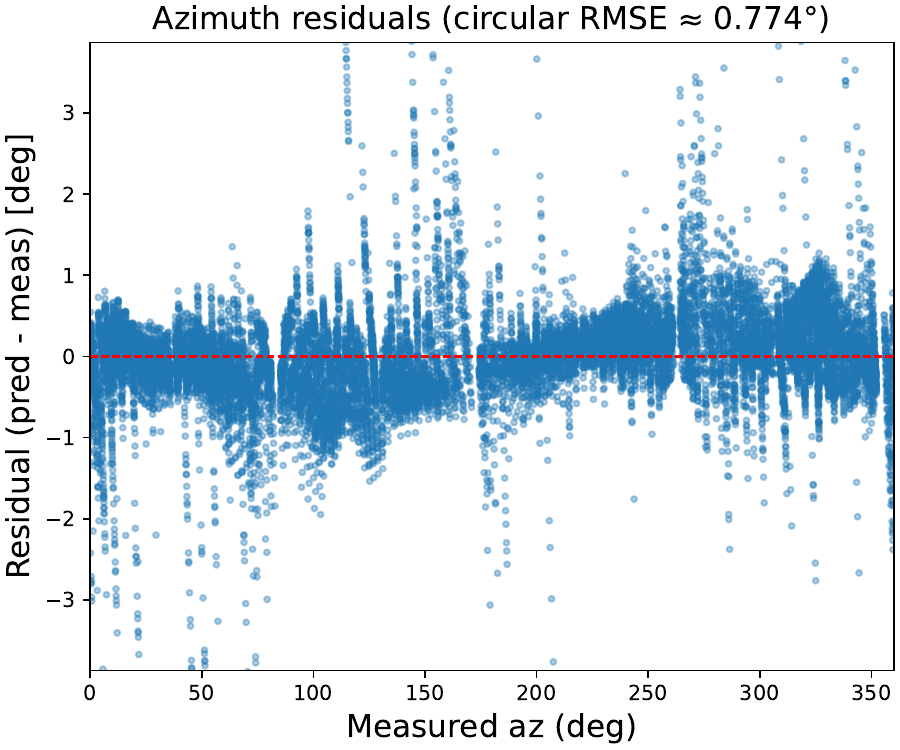}
    \caption{Azimuth fitting residual for 2019-04-24 15:39:36}
  \end{subfigure}\hfill
  \begin{subfigure}[t]{0.24\linewidth}
    \centering
    \includegraphics[width=\linewidth]{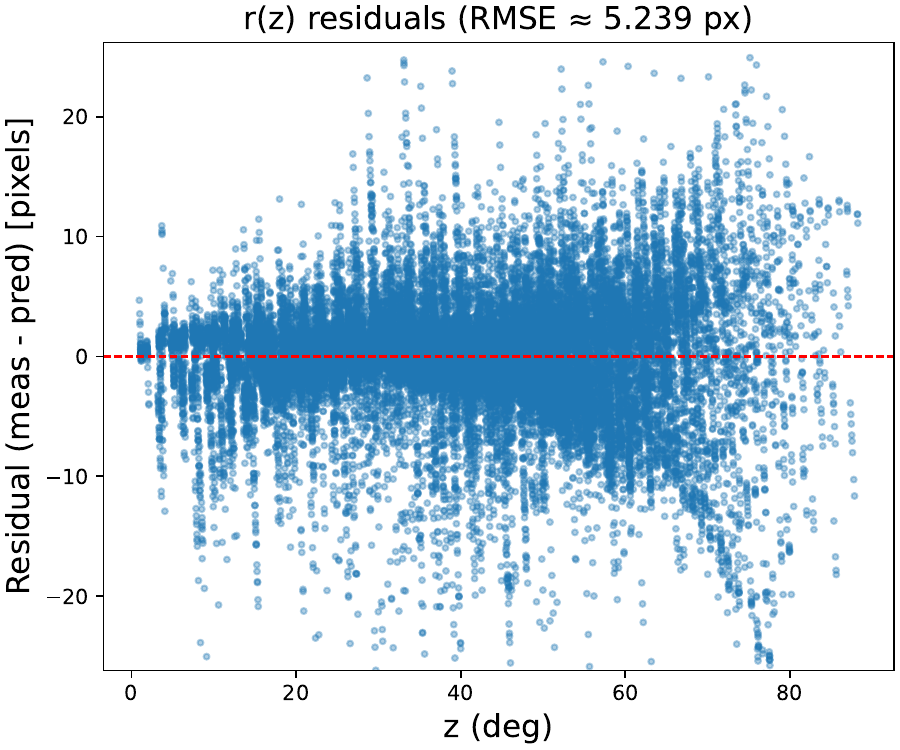}
    \caption{Radial fitting residual for 2019-06-26 18:23:18}
  \end{subfigure}\hfill
  \begin{subfigure}[t]{0.24\linewidth}
    \centering
    \includegraphics[width=\linewidth]{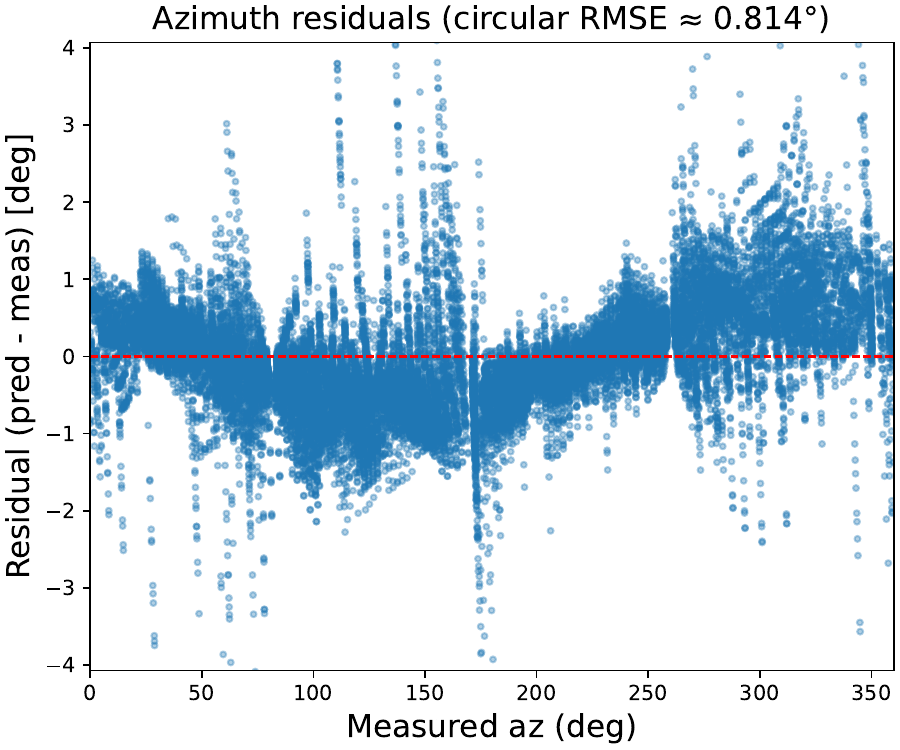}
    \caption{Azimuth fitting residual for 2019-06-26 18:23:18}
  \end{subfigure}

  \vspace{0.6em}

  \begin{subfigure}[t]{0.24\linewidth}
    \centering
    \includegraphics[width=\linewidth]{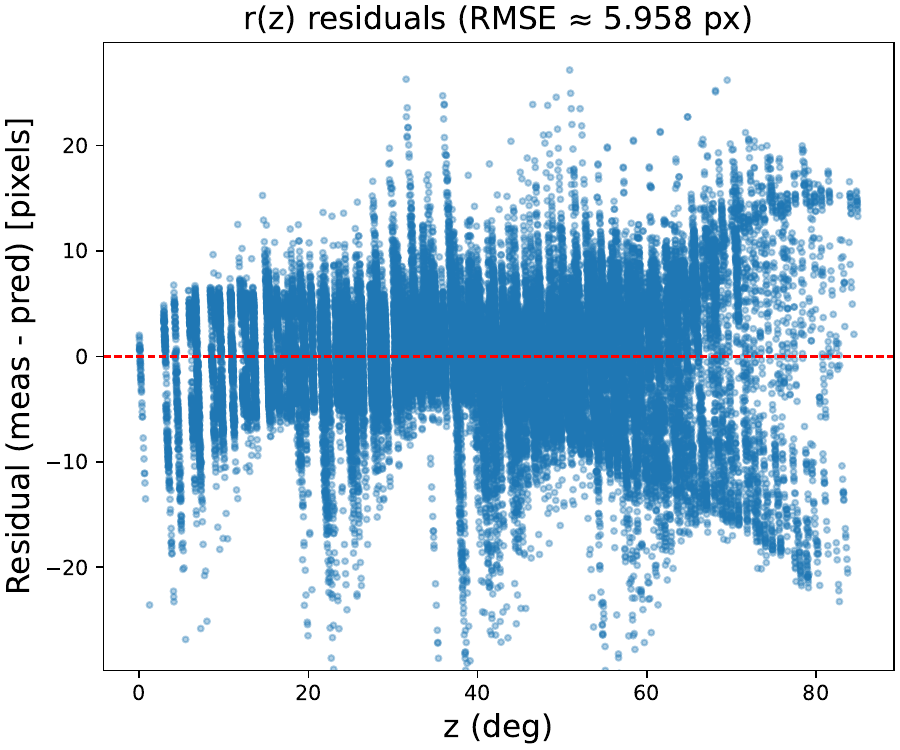}
    \caption{Radial fitting residual for 2019-07-05 11:59:14}
  \end{subfigure}\hfill
  \begin{subfigure}[t]{0.24\linewidth}
    \centering
    \includegraphics[width=\linewidth]{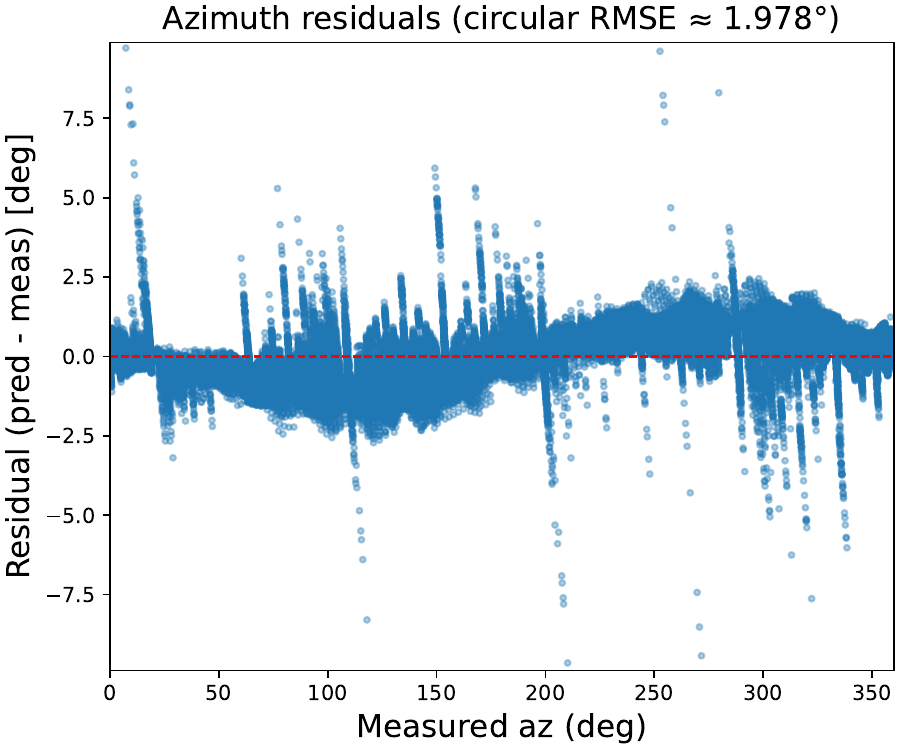}
    \caption{Azimuth fitting residual for 2019-07-05 11:59:14}
  \end{subfigure}\hfill
  \begin{subfigure}[t]{0.24\linewidth}
    \centering
    \includegraphics[width=\linewidth]{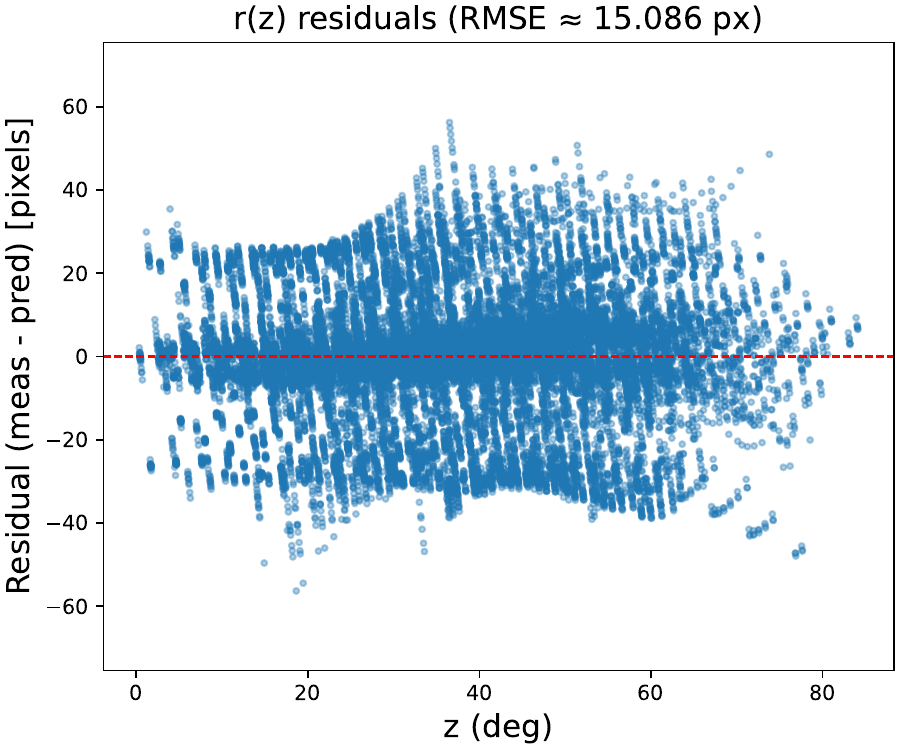}
    \caption{Radial fitting residual for 2023-09-27 18:09:48}
  \end{subfigure}\hfill
  \begin{subfigure}[t]{0.24\linewidth}
    \centering
    \includegraphics[width=\linewidth]{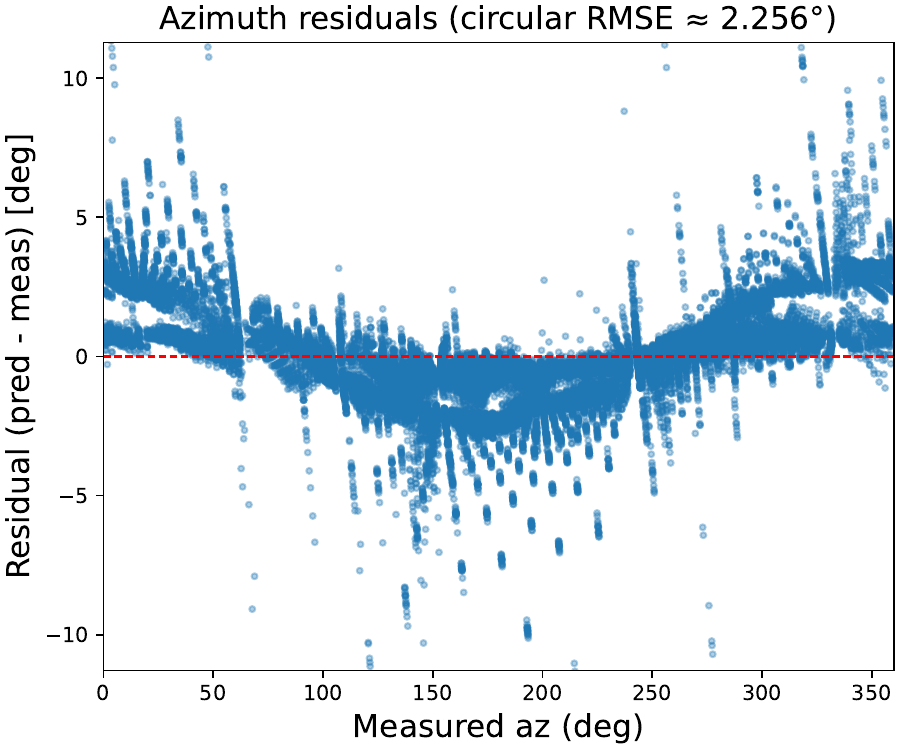}
    \caption{Azimuth fitting residual for 2023-09-27 18:09:48}
  \end{subfigure}

  \vspace{0.6em}
  \caption{Astrometric calibration fitting residual for different time slots in radial and azimuthal direction.}
  \label{fig:altazfitres}
\end{figure*}

The Altitude-Azimuth fitting map for each time slot is shown in Fig.~\ref{fig:altazfit}. 
Residual of fitting results are shown in Fig.~\ref{fig:altazfitres}.

\section{Annotation of background\label{sec:annotationdetail}}

\begin{figure*}
    \centering
    \includegraphics[width=1\linewidth]{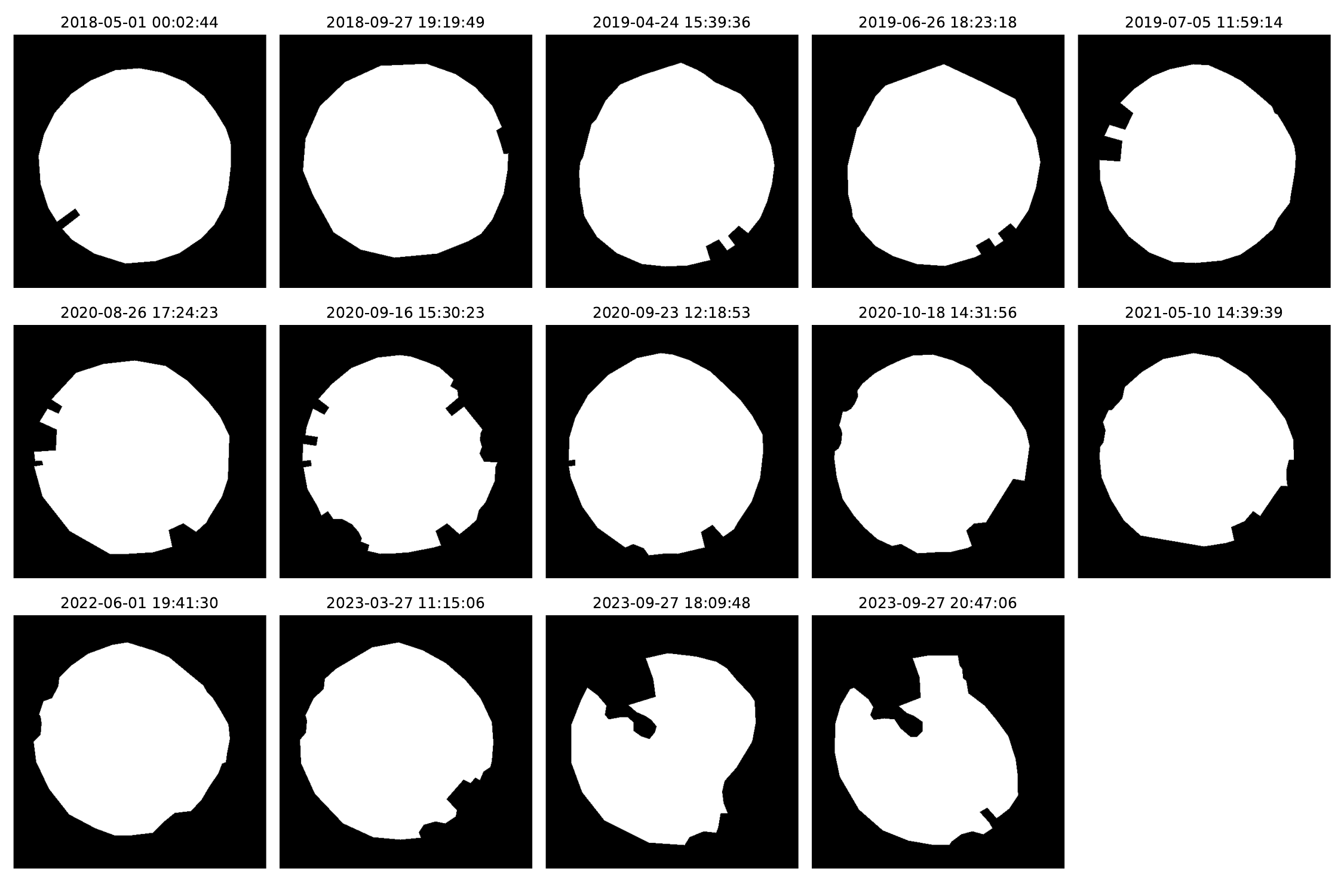}
    \caption{Annotation of all background}
    \label{fig:allbkg}
\end{figure*}

Annotation of the background as listed in Table~\ref{tab:changebkg} is shown in Fig.~\ref{fig:allbkg}.

\section{Experimental details of baseline models}
\subsection{Encoder-Decoder for cloud segmentation}
\label{sec:supp:cloudsegnet}

We use polygon annotations produced in LabelMe JSON format. For each sample, the RGB image is recovered from the embedded base64-encoded payload and converted to a three-channel array. Semantic masks are rasterized by filling the annotated polygons per class on an empty canvas that shares the image resolution. Pixels not covered by any polygon are assigned an ``ignore'' label. The class mapping is \texttt{sky}$\rightarrow$0, \texttt{cloud}$\rightarrow$1, \texttt{contamination}$\rightarrow$2, and \texttt{ignore}$\rightarrow$3. Polygons are rounded to integer coordinates and clipped to image bounds before rasterization to avoid off-by-one artifacts. 

CloudSegNet is a compact encoder--decoder CNN with two downsampling stages and two symmetric upsampling stages. A layer-wise specification of this model is shown in Table \ref{tab:segcloud}.

\begin{table*}[t]
\caption{Layer-wise specification of Encoder-Decoder (CloudSegNet) for a $3\times512\times512$ input. ``k/s/p'' denotes kernel/stride/padding. }
\label{tab:segcloud}
\centering
\small
\begin{tabular}{llllll}
\hline
Stage & Layer (in$\to$out) & Operation & k/s/p & Channel out & Size (H$\times$W) \\
\hline
Enc-1 & 3$\to$64   & Conv2d + ReLU + BN         & 3/1/1 & 64  & $512\times512$ \\
Enc-1 & 64$\to$64  & Conv2d + ReLU + BN         & 3/1/1 & 64  & $512\times512$ \\
Enc-1$\downarrow$ & -- & MaxPool2d               & 2/2/0 & 64  & $256\times256$ \\
Enc-2 & 64$\to$128 & Conv2d + ReLU + BN         & 3/1/1 & 128 & $256\times256$ \\
Enc-2 & 128$\to$128& Conv2d + ReLU + BN         & 3/1/1 & 128 & $256\times256$ \\
Enc-2$\downarrow$ & -- & MaxPool2d               & 2/2/0 & 128 & $128\times128$ \\
Dec-1$\uparrow$ & 128$\to$64 & ConvTranspose2d + ReLU & 2/2/0 & 64  & $256\times256$ \\
Dec-1 & 64$\to$64  & Conv2d + ReLU + BN         & 3/1/1 & 64  & $256\times256$ \\
Dec-2$\uparrow$ & 64$\to$32 & ConvTranspose2d + ReLU  & 2/2/0 & 32  & $512\times512$ \\
Head  & 32$\to$3   & Conv2d (logit)             & 1/1/0 & 3   & $512\times512$ \\
\hline
\end{tabular}

\end{table*}

We train with pixel-wise cross-entropy using \texttt{ignore\_index}, and uniform class weights. The optimizer is Adam with learning rate $1\mathrm{e}{-4}$. We train for up to 500 epochs and apply early stopping on validation loss with a patience of 5 epochs, retaining the checkpoint with the best validation loss.

\subsection{U-Net (CloudU-Net) for cloud segmentation}

The pre-processing steps and training parameters of U-Net is similar to encoder-decoder as above. The layer-wise specification of the applied model in this paper is shown in Table \ref{tab:unet}.

\begin{table*}[t]
\caption{Layer-wise specification of the U-Net used in our experiments (bilinear upsampling variant) for a $3\times512\times512$ input. }
\label{tab:unet}
\centering
\small
\begin{tabular}{llllll}
\hline
Stage & Layer (in$\to$out) & Operation & k/s/p & Ch.~out & Size (H$\times$W) \\
\hline
Enc-0 & 3$\to$64   & Conv2d + ReLU + BN         & 3/1/1 & 64  & $512\times512$ \\
Enc-0 & 64$\to$64  & Conv2d + ReLU + BN         & 3/1/1 & 64  & $512\times512$ \\
Down-1$\downarrow$ & -- & MaxPool2d               & 2/2/0 & 64  & $256\times256$ \\
Down-1 & 64$\to$128 & Conv2d + ReLU + BN        & 3/1/1 & 128 & $256\times256$ \\
Down-1 & 128$\to$128 & Conv2d + ReLU + BN       & 3/1/1 & 128 & $256\times256$ \\
Down-2$\downarrow$ & -- & MaxPool2d               & 2/2/0 & 128 & $128\times128$ \\
Down-2 & 128$\to$256 & Conv2d + ReLU + BN       & 3/1/1 & 256 & $128\times128$ \\
Down-2 & 256$\to$256 & Conv2d + ReLU + BN       & 3/1/1 & 256 & $128\times128$ \\
Down-3$\downarrow$ & -- & MaxPool2d               & 2/2/0 & 256 & $64\times64$ \\
Down-3 & 256$\to$512 & Conv2d + ReLU + BN       & 3/1/1 & 512 & $64\times64$ \\
Down-3 & 512$\to$512 & Conv2d + ReLU + BN       & 3/1/1 & 512 & $64\times64$ \\
Bottleneck$\downarrow$ & -- & MaxPool2d            & 2/2/0 & 512 & $32\times32$ \\
Bottleneck & 512$\to$512 & Conv2d + ReLU + BN    & 3/1/1 & 512 & $32\times32$ \\
Bottleneck & 512$\to$512 & Conv2d + ReLU + BN    & 3/1/1 & 512 & $32\times32$ \\
\hline
Up-1$\uparrow$ & 512$\to$512 & Upsample (bilinear)     & 2/--/-- & 512 & $64\times64$ \\
Up-1 & 512$+512$ & Concat (skip from Down-3)     & --      & 1024 & $64\times64$ \\
Up-1 & 1024$\to$256 & Conv2d + ReLU + BN        & 3/1/1 & 256 & $64\times64$ \\
Up-1 & 256$\to$256  & Conv2d + ReLU + BN        & 3/1/1 & 256 & $64\times64$ \\
\hline
Up-2$\uparrow$ & 256$\to$256 & Upsample (bilinear)     & 2/--/-- & 256 & $128\times128$ \\
Up-2 & 256$+256$ & Concat (skip from Down-2)     & --      & 512 & $128\times128$ \\
Up-2 & 512$\to$128 & Conv2d + ReLU + BN        & 3/1/1 & 128 & $128\times128$ \\
Up-2 & 128$\to$128 & Conv2d + ReLU + BN        & 3/1/1 & 128 & $128\times128$ \\
\hline
Up-3$\uparrow$ & 128$\to$128 & Upsample (bilinear)     & 2/--/-- & 128 & $256\times256$ \\
Up-3 & 128$+128$ & Concat (skip from Down-1)     & --      & 256 & $256\times256$ \\
Up-3 & 256$\to$64  & Conv2d + ReLU + BN        & 3/1/1 & 64  & $256\times256$ \\
Up-3 & 64$\to$64    & Conv2d + ReLU + BN        & 3/1/1 & 64  & $256\times256$ \\
\hline
Up-4$\uparrow$ & 64$\to$64   & Upsample (bilinear)     & 2/--/-- & 64  & $512\times512$ \\
Up-4 & 64$+64$   & Concat (skip from Enc-0)      & --      & 128 & $512\times512$ \\
Up-4 & 128$\to$64 & Conv2d + ReLU + BN         & 3/1/1 & 64  & $512\times512$ \\
Up-4 & 64$\to$64   & Conv2d + ReLU + BN         & 3/1/1 & 64  & $512\times512$ \\
Head  & 64$\to$3   & Conv2d (logits)             & 1/1/0 & 3   & $512\times512$ \\
\hline
\end{tabular}
\end{table*}

\subsection{SegMAN for cloud segmentation}
\label{sec:supp-segman}

The pre-processing steps and training parameters of U-Net 
is similar to encoder-decoder as above.
SegMAN is an encoder–decoder segmentation network with scalable variants (Tiny/Small/Base/Large). The parameters used in this work for these scale is exactly the same with the original paper~\cite{Fu24}. 
All variants share the same data interface and segmentation head: the model outputs dense logits that are bilinearly resized to the mask resolution when needed.
Variants differ only in encoder capacity and decoder width; we keep training/evaluation identical across variants.

\subsection{ConvLSTM for cloud nowcast}
We implement a next‐frame ConvLSTM baseline on sequences of logits with per-frame masks. Each training sample comes from \texttt{CloudLogitsDataset}, which yields a video tensor $x\in\mathbb{R}^{C\times T\times H\times W}$ together with a binary mask stack $m\in{0,1}^{T\times H\times W}$; frames are built from timestamped files and normalised per sequence by $(x-\mu)/(\sigma\cdot 6)$ before batching. For learning, we enforce $T=n_{\text{input}}{+}1$, feed the first $n_{\text{input}}$ frames to the ConvLSTM, and regress the last frame with a masked MSE on the target mask $m_{T}$; optimisation uses Adam. The configuration used in our experiments sets $n_{\text{input}}{=}2$, hidden dimensions $[64,64]$, kernel size $3$, and dropout $0$, with data sampled at 60-min intervals from user-specified time ranges. Layer-wise specification of used ConvLSTM in the paper is shown in Table~\ref{tab:convlstm-arch}.

\begin{table*}[t]
\caption{Layer-wise specification of the ConvLSTM next-frame baseline. Input is a $[B,\,C{=}3,\,T_\text{in}{=}2,\,H,\,W]$ clip; output is a single frame $[B,3,H,W]$. ``k/s/p'' denotes kernel/stride/padding.}
\label{tab:convlstm-arch}
\centering
\small
\begin{tabular}{llllll}
\hline
Stage & Layer (in$\to$out) & Operation & k/s/p & Ch.~out & Size (H$\times$W) \\
\hline
LSTM-1 & 3$\to$64   & ConvLSTM2D + Dropout        & 3/1/1 & 64  & $H\times W$ \\
LSTM-2 & 64$\to$64  & ConvLSTM2D + Dropout        & 3/1/1 & 64  & $H\times W$ \\
Head   & 64$\to$3   & Conv2d (readout)            & 1/1/0 & 3   & $H\times W$ \\
\hline
\end{tabular}
\end{table*}

\subsection{VideoGPT for cloud nowcast}
Our VideoGPT nowcaster follows a two-stage discrete latent modelling pipeline. First, a VQ-VAE encodes videos  by time-slicing, then vector-quantises features with a codebook of size $K$; training minimises masked reconstruction MSE plus standard commitment/codebook losses and tracks codebook perplexity. After training VQ-VAE, we freeze the best checkpoint and train a causal Transformer (GPT) as an autoregressive language model over the flattened VQ indices, using cross-entropy on next-token prediction. Key hyperparameters include $K{=}512$ codes for VQ-VAE and a GPT with $d_{\text{model}}{=}512$, $n_{\text{head}}{=}8$, and $6$ layers. Architecture of the VideoGPT used in this work is shown in Table~\ref{tab:videogpt-arch}.

\begin{table*}[t]
\caption{Architecture of the VideoGPT nowcaster (VQ-VAE + GPT). For a $[B,3,T,64,64]$ input, the encoder downsamples to $[B,256,T,8,8]$, which is vector-quantised (codebook size $K$). Tokens are modeled autoregressively by a Transformer and decoded back to frames.}
\label{tab:videogpt-arch}
\centering
\small
\begin{tabular}{llllll}
\hline
\multicolumn{6}{l}{\textbf{VQ-VAE Encoder}} \\
\hline
Enc-1 & 3$\to$64    & Conv2d + BN + ReLU + ResidualBlock & 4/2/1 & 64  & $32\times32$ \\
Enc-2 & 64$\to$128  & Conv2d + BN + ReLU + ResidualBlock & 4/2/1 & 128 & $16\times16$ \\
Enc-3 & 128$\to$256 & Conv2d + BN + ReLU + ResidualBlock & 4/2/1 & 256 & $8\times8$ \\
Bottleneck & 256$\to$256 & Conv2d + BN + ReLU            & 3/1/1 & 256 & $8\times8$ \\
\hline
\multicolumn{6}{l}{\textbf{Vector Quantiser}} \\
\hline
VQ & 256$\to K$ & VectorQuantizer ($K{=}512$, $D{=}256$) & -- & $K$ & $T\times8\times8$ \\
\hline
\multicolumn{6}{l}{\textbf{VQ-VAE Decoder}} \\
\hline
Dec-3$\uparrow$ & 256$\to$128 & ConvTranspose2d + BN + ReLU + ResidualBlock & 4/2/1 & 128 & $16\times16$ \\
Dec-2$\uparrow$ & 128$\to$64  & ConvTranspose2d + BN + ReLU + ResidualBlock & 4/2/1 & 64  & $32\times32$ \\
Dec-1$\uparrow$ & 64$\to$64   & ConvTranspose2d + BN + ReLU + ResidualBlock & 4/2/1 & 64  & $64\times64$ \\
Head            & 64$\to$3    & Conv2d + Tanh                                & 3/1/1 & 3   & $64\times64$ \\
\hline
\multicolumn{6}{l}{\textbf{GPT (autoregressive over tokens)}} \\
\hline
TokEmb & $K\to d_\text{model}$ & Embedding                                   & --    & 512 & -- \\
PosEmb & --          & Learned positional embedding (length 20{,}000)   & --    & 512 & -- \\
Transf & 512$\to$512 & \#Layers$=6$, \#Heads$=8$ (TransformerEncoderLayer) & -- & 512 & -- \\
LN     & 512$\to$512 & LayerNorm                                         & --    & 512 & -- \\
Head   & 512$\to K$  & Linear (projection to vocab)                       & --    & $K$ & -- \\
\hline
\end{tabular}
\end{table*}

\section{Per-class experimental results\label{sec:perclassres}}
 The per-class nowcasting results are shown in Table~\ref{tab:perclassnowcast}; The per-class cloud segmentation results are shown in Table~\ref{tab:perclassseg}.

\begin{table*}[ht]
\centering
\caption{Per-class metrics for nowcasting\label{tab:perclassnowcast}}
\begin{tabular}{llccc}
\hline
Class & Baseline & Precision & Recall & F1-score \\
\hline
\multirow{6}{*}{Sky}  & Trivial & 0.914 & \textbf{0.913} & \textbf{0.914} \\
 & Optica lFlow & 0.913 & 0.913 & 0.913 \\
& ConvLSTM & \textbf{0.922} & 0.905 & 0.913 \\
 & VideoGPT-1 & 0.920 & 0.882 & 0.900 \\
 & VideoGPT-2 & 0.915 & 0.887 & 0.901 \\
 & VideoGPT-7 & 0.919 & 0.883 & 0.900 \\
\hline
\multirow{6}{*}{Cloud} & Trivial & 0.876 & 0.878 & 0.877 \\
 & Optical Flow & \textbf{0.878} & 0.878 & 0.878 \\
& ConvLSTM & 0.860 & \textbf{0.907} & \textbf{0.883} \\
 & VideoGPT-1 & 0.833 & 0.904 & 0.867 \\
 & VideoGPT-2 & 0.841 & 0.897 & 0.868 \\
 & VideoGPT-7 & 0.830 & 0.903 & 0.865 \\
\hline
\multirow{6}{*}{Contamination}  & Trivial & 0.787 & 0.782 & 0.785 \\
 & Optical Flow & 0.784 & \textbf{0.788} & \textbf{0.786} \\
& ConvLSTM & \textbf{0.831} & 0.734 & 0.779 \\
 & VideoGPT-1 & 0.768 & 0.685 & 0.724 \\
 & VideoGPT-2 & 0.763 & 0.690 & 0.724 \\
 & VideoGPT-7 & 0.774 & 0.668 & 0.717 \\
\hline
\end{tabular}
\end{table*}

\begin{table*}[ht]
\centering
\caption{Per-class metrics for cloud segmentation(median with 16th-83rd percentiles)\label{tab:perclassseg}}
\begin{tabular}{llccc}
\hline
Class & Baseline & Precision & Recall & F1-score \\
\hline
\multirow{10}{*}{Sky} & DINOv3 local & $0.920_{-0.022}^{+0.028}$ & $0.945_{-0.009}^{+0.018}$ & $0.936_{-0.015}^{+0.006}$ \\
 & DINOv3 local + CLS & $0.917_{-0.025}^{+0.037}$ & $0.941_{-0.014}^{+0.019}$ & $0.932_{-0.013}^{+0.012}$ \\
 & DINOv3 local + CLS + register & $0.913_{-0.031}^{+0.038}$ & $0.942_{-0.025}^{+0.018}$ & $0.926_{-0.012}^{+0.015}$ \\
 & DINOv3 local + mean(CLS + register) & $0.921_{-0.019}^{+0.031}$ & $0.944_{-0.019}^{+0.018}$ & $0.936_{-0.016}^{+0.009}$ \\
 & Encoder-Decoder & $0.764_{-0.027}^{+0.061}$ & $0.868_{-0.034}^{+0.056}$ & $0.816_{-0.009}^{+0.021}$ \\
 & U-Net & $0.869_{-0.039}^{+0.023}$ & $0.912_{-0.047}^{+0.025}$ & $0.882_{-0.016}^{+0.017}$ \\
& SegMAN Tiny & $0.881_{-0.019}^{+0.027}$ & $0.926_{-0.027}^{+0.022}$ & $0.906_{-0.017}^{+0.008}$ \\
 & SegMAN Small & $0.871_{-0.031}^{+0.026}$ & $0.929_{-0.019}^{+0.031}$ & $0.904_{-0.010}^{+0.013}$ \\
 & SegMAN Base & $0.874_{-0.027}^{+0.017}$ & $0.927_{-0.017}^{+0.025}$ & $0.901_{-0.012}^{+0.010}$ \\
 & SegMAN Large & $0.854_{-0.014}^{+0.025}$ & $0.931_{-0.026}^{+0.021}$ & $0.899_{-0.021}^{+0.005}$ \\

\hline
\multirow{10}{*}{Cloud} & DINOv3 local & $0.961_{-0.018}^{+0.013}$ & $0.966_{-0.011}^{+0.011}$ & $0.961_{-0.010}^{+0.013}$ \\
 & DINOv3 local + CLS & $0.965_{-0.024}^{+0.006}$ & $0.969_{-0.016}^{+0.007}$ & $0.960_{-0.012}^{+0.012}$ \\
 & DINOv3 local + CLS + register & $0.954_{-0.016}^{+0.019}$ & $0.965_{-0.011}^{+0.013}$ & $0.960_{-0.013}^{+0.009}$ \\
 & DINOv3 local + mean(CLS + register) & $0.966_{-0.021}^{+0.010}$ & $0.970_{-0.017}^{+0.007}$ & $0.962_{-0.010}^{+0.013}$ \\
 & Encoder-Decoder & $0.823_{-0.034}^{+0.052}$ & $0.823_{-0.047}^{+0.050}$ & $0.822_{-0.035}^{+0.028}$ \\
 & U-Net & $0.863_{-0.062}^{+0.035}$ & $0.921_{-0.031}^{+0.017}$ & $0.880_{-0.036}^{+0.020}$ \\
 & SegMAN Tiny & $0.902_{-0.027}^{+0.031}$ & $0.925_{-0.042}^{+0.018}$ & $0.910_{-0.019}^{+0.017}$ \\
 & SegMAN Small & $0.905_{-0.029}^{+0.031}$ & $0.912_{-0.047}^{+0.036}$ & $0.904_{-0.015}^{+0.020}$ \\
 & SegMAN Base & $0.900_{-0.029}^{+0.028}$ & $0.921_{-0.038}^{+0.021}$ & $0.913_{-0.024}^{+0.008}$ \\
 & SegMAN Large & $0.899_{-0.034}^{+0.025}$ & $0.900_{-0.024}^{+0.032}$ & $0.899_{-0.024}^{+0.018}$ \\

\hline
\multirow{10}{*}{Contamination} & DINOv3 local & $0.808_{-0.092}^{+0.056}$ & $0.686_{-0.121}^{+0.080}$ & $0.726_{-0.055}^{+0.059}$ \\
 & DINOv3 local + CLS & $0.783_{-0.057}^{+0.068}$ & $0.644_{-0.068}^{+0.135}$ & $0.715_{-0.050}^{+0.053}$ \\
 & DINOv3 local + CLS + register & $0.806_{-0.084}^{+0.060}$ & $0.630_{-0.113}^{+0.121}$ & $0.707_{-0.059}^{+0.067}$ \\
 & DINOv3 local + mean(CLS + register) & $0.793_{-0.087}^{+0.073}$ & $0.675_{-0.114}^{+0.126}$ & $0.715_{-0.042}^{+0.047}$ \\
 & Encoder-Decoder & $0.730_{-0.124}^{+0.108}$ & $0.285_{-0.071}^{+0.077}$ & $0.396_{-0.056}^{+0.077}$ \\
 & U-Net & $0.839_{-0.278}^{+0.055}$ & $0.280_{-0.089}^{+0.076}$ & $0.415_{-0.130}^{+0.072}$ \\
& SegMAN Tiny & $0.637_{-0.126}^{+0.113}$ & $0.437_{-0.092}^{+0.112}$ & $0.496_{-0.052}^{+0.100}$ \\
 & SegMAN Small & $0.678_{-0.153}^{+0.117}$ & $0.383_{-0.077}^{+0.121}$ & $0.489_{-0.057}^{+0.043}$ \\
 & SegMAN Base & $0.683_{-0.065}^{+0.069}$ & $0.376_{-0.105}^{+0.089}$ & $0.483_{-0.095}^{+0.057}$ \\
 & SegMAN Large & $0.654_{-0.112}^{+0.145}$ & $0.359_{-0.087}^{+0.066}$ & $0.464_{-0.065}^{+0.045}$ \\
\hline
\end{tabular}
\end{table*}

\end{document}